%% file: Priyanka_paperI.tex
\documentclass[twocolumn,trackchanges]{aastex63}

\usepackage{graphicx,subfigure}
\usepackage[utf8]{inputenc}
\accepted{for publication in ApJ}

\usepackage{longtable}
\usepackage{booktabs}
\usepackage{multirow}
\usepackage{savesym}
\savesymbol{tablenum}
\usepackage{siunitx}
\restoresymbol{SIX}{tablenum}
\usepackage{natbib}
\usepackage{amsmath}
\graphicspath{{./}{figures/}}
\input{macros.tex}

\usepackage{xcolor}
\usepackage{appendix}
 \begin{document}

\title{X-ray spectroscopy in the microcalorimeter era I: Effects of  Fe XXIV  Resonance Auger Destruction on Fe XXV K$\alpha$ spectra}
\author[0000-0002-4469-2518]{P. Chakraborty}
\affiliation{University of Kentucky \\
Lexington, KY, USA}
\author[0000-0003-4503-6333]{G. J. Ferland}
\affiliation{University of Kentucky \\
Lexington, KY, USA}
\author[0000-0002-8823-0606]{M. Chatzikos}
\affiliation{University of Kentucky \\
Lexington, KY, USA}
\author[0000-0002-2915-3612]{F. Guzm\'an}
\affiliation{University of Kentucky \\
Lexington, KY, USA}
\author{Y. Su}
\affiliation{University of Kentucky \\
Lexington, KY, USA}

\begin{abstract}

We discuss the importance of  Fe$^{23+}$ in determining the line intensities of the Fe XXV K$\alpha$ complex in an optically thick cloud, and investigate the prediction of Liedahl (2005) on Resonance Auger Destruction (RAD) with CLOUDY. Although initially motivated by the Perseus cluster, our calculations are extended to the wide range of column densities encountered in astronomy. A Fe XXV line photon can change/lose its identity upon absorption by three-electron iron as a result of ``line interlocking".  This may lead to the autoionization of the absorbing ion, ultimately destroying the Fe XXV K$\alpha$ photon by RAD. Out of the four members in the  Fe XXV K$\alpha$ complex, a significant fraction of the x line photons is absorbed by Fe$^{23+}$ and destroyed, causing the x line intensity to decrease. For example, at a hydrogen column density of 10$^{25}$ cm$^{-2}$, $\sim$ 32\% of x photons are destroyed due to RAD while w is mostly unaffected. The line intensity of y is slightly ($\leq$2\%) reduced. z is not directly affected by RAD, but the contrasting behavior between z and x line intensities points towards the possible conversion of a tiny fraction ($\sim$ 2\%) of x photons into z photons.  The change in line intensities due to Electron Scattering Escape (ESE) off fast thermal electrons is also discussed.

  
\end{abstract}


\section{Introduction}
The Fe XXV K$\alpha$ complex has been a subject of interest in the X-ray community for decades. Observations of the Perseus cluster by Ariel 5 detected an emission feature near 7 keV caused by transitions from Fe XXV and Fe XXVI \citep{1976MNRAS.175P..29M}. The  Fe XXV K$\alpha$ complex was later detected at 6.7 keV by XMM-Newton \citep {2004MNRAS.347...29C}. Recently, the Soft X-ray Spectrometer  \citep[SXS,][]{2016SPIE.9905E..0VK} on board \textit{Hitomi} resolved the complex into four components – the resonance (w), intercombination (x,y), and forbidden (z) lines \citep{2016Natur.535..117H}. 

In the higher column density limit, two atomic processes contribute to the change in intensity of line photons 
in the Fe XXV K$\alpha$ complex. First, line photons are absorbed by satellites of three- electron iron and He-like chromium due to line 
interlocking and lose their identities (see section \ref{Line interlocking}, \ref{autoionization}). Second, Lyman 
line photons are absorbed by Fe XXV itself following the emission of different
line photons (Case A to B transition), as discussed in the second paper of this series \citep[][hereafter paper II]{Chakraborty2_2020}. 
However, it is difficult to 
decouple these two processes and study them independently. We focus this paper on the importance of line interlocking 
in deciding Fe XXV K$\alpha$ line intensities using the spectral synthesis code CLOUDY \citep{2017RMxAA..53..385F}, with
occasional references to the Case A to B transition.

We also explore the effects of Resonance Auger Destruction (RAD),
originally introduced by \citet{2005AIPC..774...99L}, and find it very important to explain the intensities of selective lines in the Fe XXV K$\alpha$ complex in the high-column-density limit. Although we do all
our calculations using the physical parameters for the Perseus core, which is optically thin for x, y, and z,
RAD only becomes important in the high-column-density limit. This is a spectral analysis in the column density parameter space, 
extended up to the hydrogen column density N$_{\rm{H}}$=10$^{25}$ cm$^{-2}$, typically encountered in astronomy. Although such high column densities do not occur in the Perseus intracluster medium, they do happen in other environments, such as accretion disks, Compton thick regions, Fe K$\alpha$ fluorescent emission line in active galactic nuclei, and some Seyfert galaxies \citep{10.1093/mnras/280.3.823, 2002A&A...387...76B,  doi:10.1111/j.1365-2966.2005.08661.x, 2011MNRAS.412..277Y, 2012MNRAS.423.3360Y, 2013MNRAS.436.1615M, 2019ApJ...885...62T} which CLOUDY can model. We document the full physical treatment for future reference.


In addition, at hydrogen column densities higher than N$_{\rm{H}}$=10$^{23}$ cm$^{-2}$, electron scattering opacity starts to become important. Line photons scatter off thermal electrons and become heavily Doppler shifted, leading to one-scattering escape. We call this process Electron Scattering Escape (ESE). This causes a deficit in the line intensity at the wavelength of the scattered photons. We elaborate on this process in section \ref{ESE}. 

The organization of this paper is as follows. Section \ref{Atomic Processes} lists the various atomic data sources used in our calculations. Section \ref{Simulation Parameters} discusses the parameters used for our simulations with CLOUDY. Section \ref{Line interlocking} discusses the physics of line interlocking. Section \ref{autoionization} demonstrates autoionization following the absorption of Fe XXV K$\alpha$ line photons. Section \ref{Spectroscopic evidence} discusses the spectroscopic evidence of RAD in selective Fe XXV K$\alpha$ line photons. Section \ref{ESE} describes the change in Fe XXV K$\alpha$ line intensities due to Electron Scattering Escape. Section \ref{Summary} summarizes our results.

\section{Atomic data}\label{Atomic Processes}
This section describes the various atomic data sources we use in our spectral modeling.
Our calculations are very wavelength sensitive and require a precise atomic dataset 
for the accurate prediction of spectral behavior in the high-column-density limit. 
Atomic data sources for He-like iron (Fe$^{24+}$) are discussed in paper II.
Atomic data sources for the two ions contributing to line interlocking with Fe$^{24+}$: Fe$^{23+}$ and Cr$^{22+}$
(for hydrogen column densities higher than 10$^{23}$ cm$^{-2}$) are discussed below.

\subsection{Energy Levels}
The previous version of CLOUDY \citep{2017RMxAA..53..385F} used Fe$^{23+}$ and Cr$^{22+}$ energy levels calculated using the
 code Autostructure, as described in \citet{2005MNRAS.360..458B}. We replace these with CHIANTI version 9.0 \citep{2019ApJS..241...22D,1997A&AS..125..149D},
 which uses a more recent Autostructure calculation \citep{2011CoPhC.182.1528B}.
 The process of ``line interlocking" described in section \ref{Line interlocking} is very sensitive to the line wavelengths, 
 as even a slight variation in the energy levels can significantly change the line-center optical depths. We find that the uncertainty in Fe$^{23+}$ is particularly important for our analysis. See section \ref{Summary} for a brief discussion on how uncertainties
 in energy levels or velocity fields can alter the nature of the spectra in high column densities.

\begin{table*}
\centering{
  \caption{\label{t:E_Fe XXIV}Energy level configurations, labels, corresponding wavelengths, downward transition probabilities (A$_{u,l}$), and autoionization rates (A$_{a}$) (if applicable)  for transitions from the ground in Fe$^{24+}$  generating Fe XXV K$\alpha$ complex, and in Fe$^{23+}$, Fe$^{22+}$, and Cr$^{22+}$ in the proximity of the Fe XXV K$\alpha$ complex. }}
  
\begin{tabular}{llllrrr}
\hline
Ion& Configuration & Label & Wavelength(\AA)& \hspace{5mm}A$_{u,l}$(s$^{-1}$) & \hspace{5mm}A$_{a}$(s$^{-1}$)\\
\hline
&$1s.2p \hspace{2mm} \rm{^{1}P_{1}} $ &\hspace{2mm} w &\hspace{5mm}1.8504 &  4.54e+14 &  -  \\  
Fe$^{24+}$&$1s.2p\hspace{2mm} \rm{^{3}P_{2}}$ &\hspace{2mm} x &\hspace{5mm}1.8554 & 6.30e+09 &  -  \\  
&$1s.2p\hspace{2mm} \rm{^{3}P_{1}}$ &\hspace{2mm} y &\hspace{5mm}1.8595 &  4.11e+13 &   -    \\
&$1s.2s\hspace{2mm} \rm{^{3}S_{1}}$ &\hspace{2mm} z &\hspace{5mm}1.8682 & 2.15e+08 &  -  \\   
\hline
&$1s.2s.2p \hspace{2mm} \rm{^{2}P_{3/2}} $&\hspace{2mm} -&\hspace{5mm}1.8563 &  3.420e+12 &  1.09e+14  \\  
Fe$^{23+}$&$1s.2s.2p\hspace{2mm} \rm{^{2}P_{1/2}}$ &\hspace{2mm} -&\hspace{5mm}1.8571 &  1.900e+14 &  7.58e+13  \\  
&$1s.2s.2p\hspace{2mm} \rm{^{2}P_{3/2}}$ &\hspace{2mm} -&\hspace{5mm}1.8610 &  4.680e+14&   2.90e+09    \\
&$1s.2s.2p\hspace{2mm} \rm{^{2}P_{1/2}}$ &\hspace{2mm} -&\hspace{5mm}1.8636 &  2.920e+14&  3.38e+13  \\   
\hline
Cr$^{22+}$&$1s.3p\hspace{9mm} \rm{^{1}P_{1}}$ &\hspace{2mm} -&\hspace{5mm}1.8558 &  8.970e+13 &-   \\  
\hline
Fe$^{22+}$&$1s.2s2.2p\hspace{2mm} \rm{^{1}P_{1}}$ &\hspace{2mm} -&\hspace{5mm}1.8704 &  4.210e+14 &  5.56e+12  \\  
\hline
\end{tabular}
 \end{table*}

\subsection{Transition Probabilities}

For transitions between autoionizing levels and ground in Fe$^{23+}$, we use transition probabilities from CHIANTI version 9.0, which are taken from Autostructure calculations. Table \ref{t:E_Fe XXIV} gives the list of transition probabilities used in our calculation near the energy range of the Fe XXV K$\alpha$ complex.
We found a swap between two Fe XXIV lines in the NIST atomic database \citep{2018APS..DMPM01004K}, which can significantly alter the line-center optical depth in x, as well as the calculations for RAD. This issue is further addressed in section \ref{Summary}.

\subsection{Autoionization rates}

We use autoionization rates of autoionizing energy levels in Fe$^{23+}$ from Autostructure \citep{2011CoPhC.182.1528B}.
The list of autoionization rates for the energy levels near  Fe XXV K$\alpha$ energies is given in  Table \ref{t:E_Fe XXIV}.


\section{Simulation Parameters}\label{Simulation Parameters}
For our simulation with CLOUDY, we use the temperature, metal abundance, and turbulence in Perseus core. 
Although the hydrogen column density reported by \citet{2018PASJ...70...12H} ($ N_{\rm H} \sim  1.88\times 10^{21} \pscm$) 
is too small in Perseus to show the effect of RAD,  we extend our parameter space up to a hydrogen column density $ N_{\rm H} =  10^{25} \pscm$. This is to document models of optically thick environments to which CLOUDY can be applied.
We choose the temperature  and Fe  abundance in the  interval 4.05$^{+0.01}_{-0.01}$ keV and 0.65$^{+0.01}_{-0.01}$ 
of solar, respectively, following the \textit{Hitomi} observations \citep{2018PASJ...70...10H} from a broad-band fit in the energy range 1.8-20.0 keV.
We assume an average hydrogen 
density of 0.03 cm$^{-3}$ in our region of interest, and a chromium to iron abundance ratio $\sim$ 0.15.
Although the \citet{2018PASJ...70...10H} reported a slight difference in turbulent broadening between 
w (159–167 km/s) and x, y, z (136–150 km/s) in the outer core of Perseus,
we use a turbulent velocity of 150 km/s in our simulation for simplicity. 

\section{Loss of identity via ``Line interlocking"}\label{Line interlocking}

\begin{figure*}
\gridline{\fig{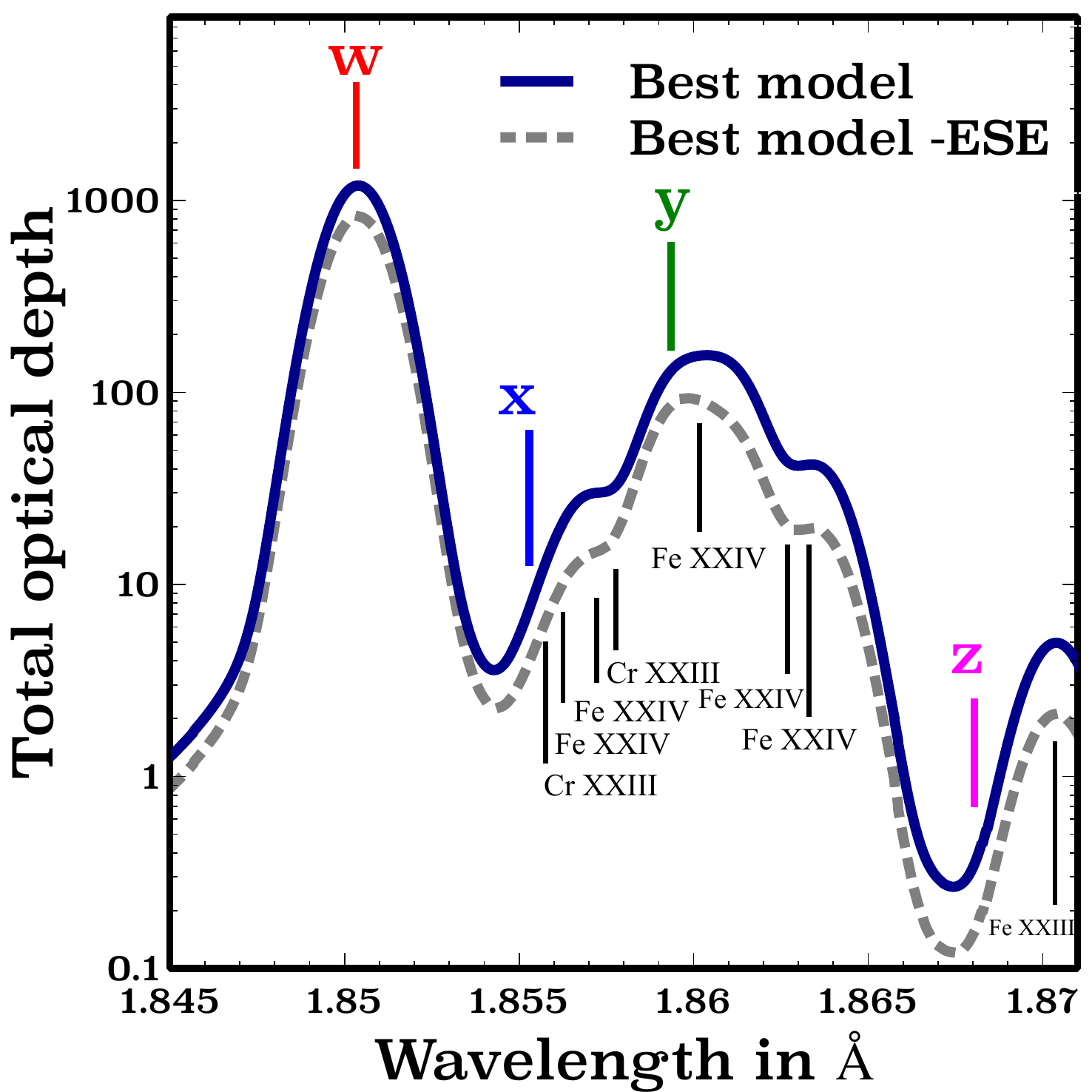}{0.4\textwidth}{(a)}
          \fig{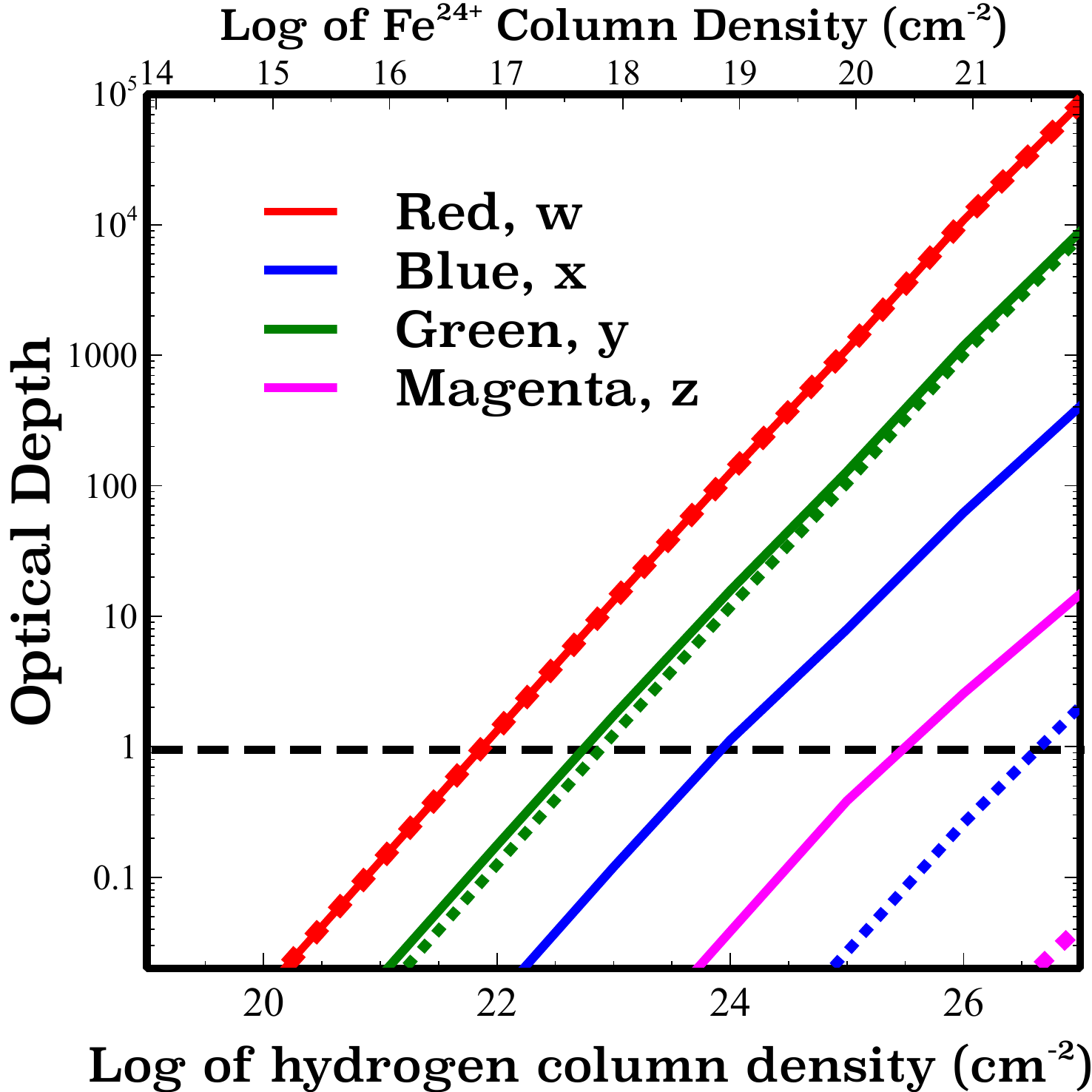}{0.42\textwidth}{(b)}
          }

\caption{a) Total optical depth for a cloud with a hydrogen column density of $10^{25}$ cm$^{-2} $ 
plotted as a function of wavelength in \si{\angstrom}. Solid and dashed lines show our best model
and the best model without Electron Scattering Escape (ESE), respectively.
The vertical lines above the blue curve mark the wavelengths of w, x, y, and z.
Individual lines with a single-line optical depth $\geq 1$, are marked with vertical lines below the grey dashed curve.
b) Variation of total line-center optical depth, and single-line optical depth for x, y, z, and w 
with hydrogen column density for our best model. Solid lines show the total line-center optical depths, 
and dashed lines show single-line optical depths for these lines. 
Slower than linear increase in the total line-center optical depths in x, y, and z photons in the high-column-density limit is due to the decline in the Fe$^{23+}$ and Fe$^{22+}$ concentration following autoionization and RAD.
\label{f:Lineinterlockig}}
\end{figure*}

The total optical depth of the cloud determines the radiative transfer effects in line photons like absorption by ions of the same/ different element, or the Case A to B transfer (see paper II). A photon within a line simply scatters off the total opacity that is present in the gas. 
It has no knowledge of which transition or species created that opacity. 
This scattering produces “line interlocking” where a photon can change its identity.  
This process is described, for instance, in \citet{1985ApJ...291..464E} and \citet{1985ApJ...299..752N}.

 CLOUDY handles line overlap with a “fine opacity” grid that is described by \citet{2005ApJ...624..794S}. The left panel of Figure \ref{f:Lineinterlockig} shows this fine opacity grid in the spectral region near Fe XXV K$\alpha$
for $ N_{\rm H} = 10^{25} \pscm$, the upper limit of column density used in our calculations. 
For w, the total line-center optical depth solely comes from its single-line optical depth. 
But x, y, and z have contributions from other lines.  
The dominant contributions to the total optical depth at the center of  x  are from Fe XXIV ($\lambda$=1.8571\AA) and 
Cr XXIII ($\lambda$=1.8558\AA) satellites, and z is from Fe XXIII ($\lambda$=1.8704\AA). 
y has a small contribution from the Fe XXIV satellite ($\lambda$= 1.8610\AA).
Electron Scattering Escape (ESE) also becomes important at such a high column density.
The Figure shows the variation of the total line-center optical depth with wavelength for our best model and a model without ESE  with solid and dashed lines, respectively.

The right panel of Figure \ref{f:Lineinterlockig} shows the contrast between the single-line and total 
line-center optical depths of x, y, z, w 
for our best model, and their variation with column density.
The single-line optical depths in x and z significantly differ from that of their line-center optical depths.
This difference is a result of absorption of x and z photons by Fe$^{23+}$ and Cr$^{22+}$, and Fe$^{22+}$ respectively at the wavelengths mentioned in the previous paragraph. The total
line-center optical depth in y has a relatively smaller deviation from its single-line optical depth.
This happens due to absorption by 
Fe$^{23+}$. There is an overlap between the two optical depths in w as expected, with zero absorption by other ions.

The total line-center optical depth  shows a linear increase with column density for w. However, x, y, and z show
a slightly slower than linear increase in their total line-center optical depth with column density 
in the high-column-density limit.
Such deviations from linearity are caused by a decline in the Fe$^{23+}$  and Fe$^{22+}$ concentration through RAD \citep{2005AIPC..774...99L}. See the next section for further discussion.



\section{Autoionization following absorption}\label{autoionization}

\begin{figure}
\subfigure{\includegraphics[width = 3in]{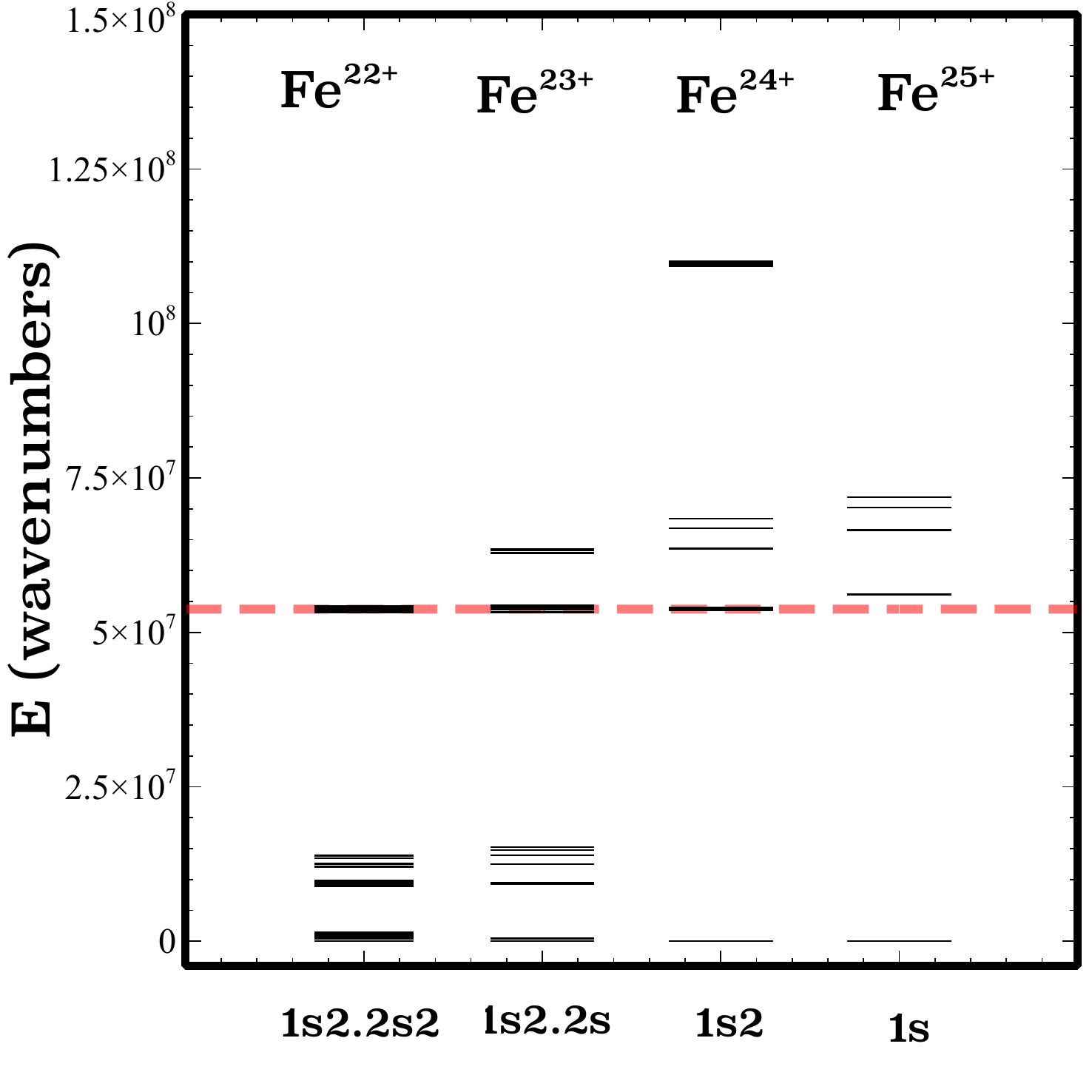}}\hspace{2cm}
\subfigure{\includegraphics[width = 3in]{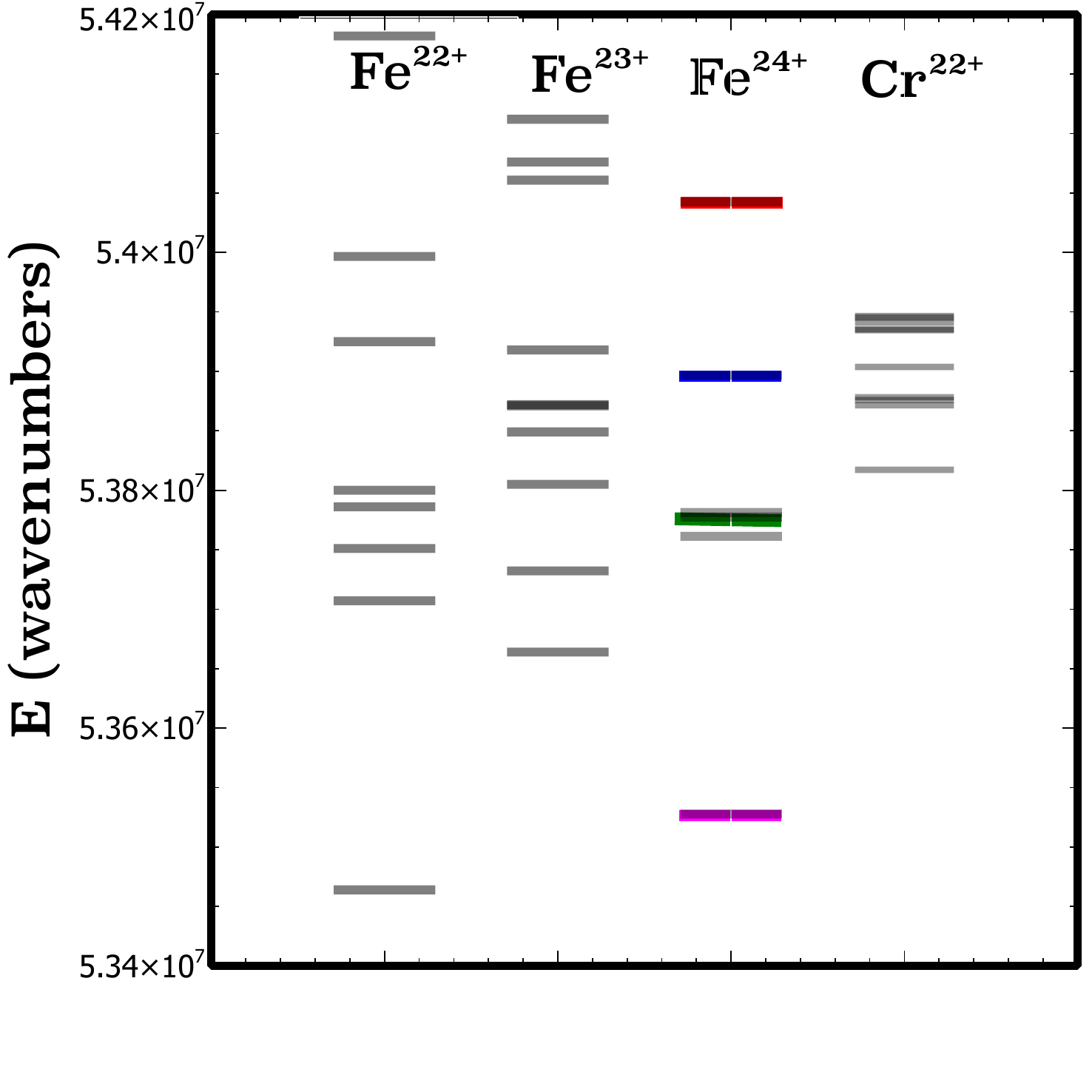}} 
\caption{Top: Partial energy level diagram of one-, two-, three-, and four-electron iron. 
The red dashed line marks the energy of the Fe XXV K$\alpha$ complex.
      Bottom: Enlarged energy diagram of Fe$^{22+}$, Fe$^{23+}$, Fe$^{24+}$, and  Cr$^{22+}$ near Fe XXV K$\alpha$ complex. Red, blue, green, and magenta lines mark the upper energy level of w, x, y, and z transition.}
\label{fig:FeXXIII_FeXXIV_FeXXV_FeXXVI}
\end{figure}

\begin{figure}[h!]
\centering
\includegraphics[scale=0.5]{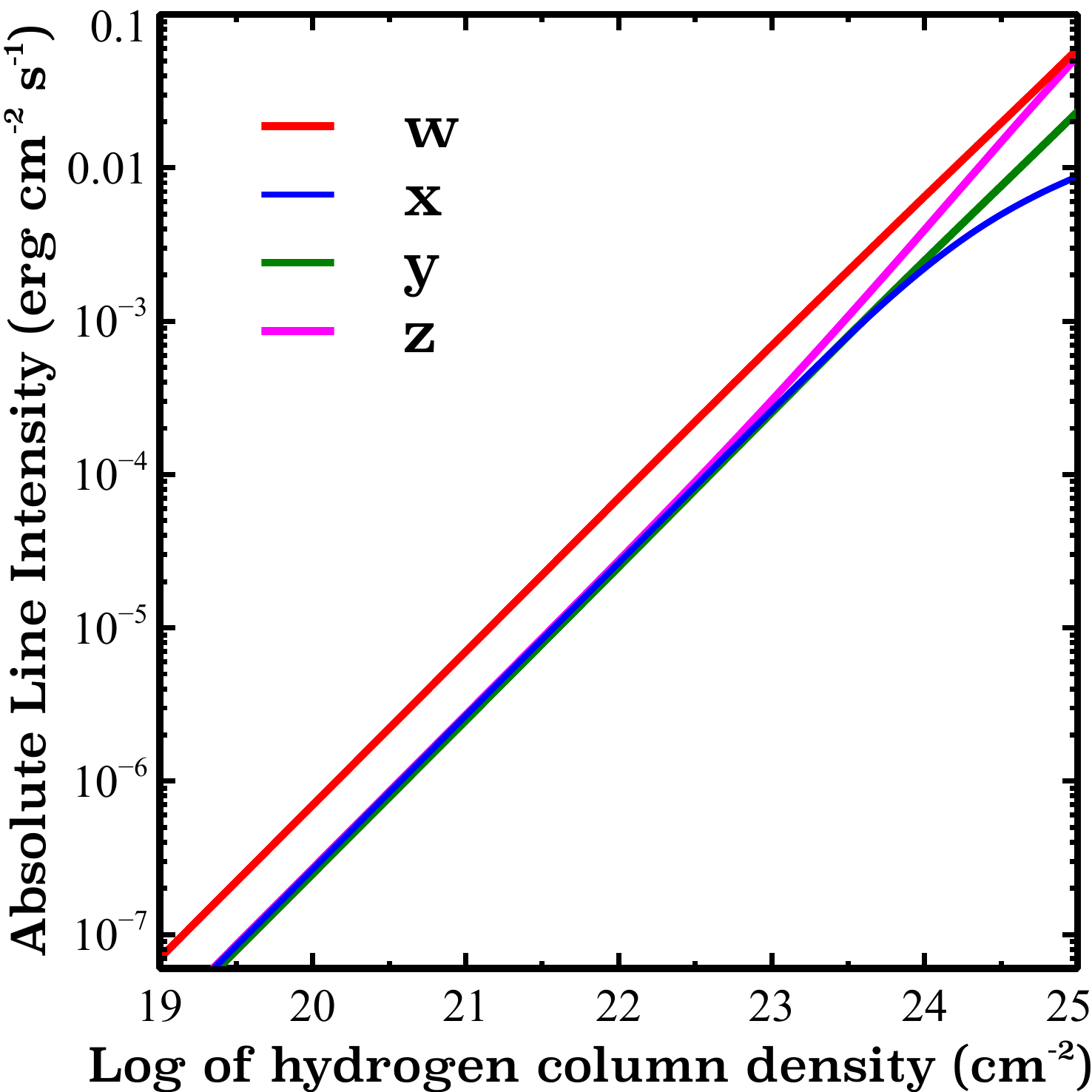}
\caption{ Absolute line intensities for the members of the Fe XXV K$\alpha$ complex with respect to hydrogen column density.  }
\label{fig:absolute}
\end{figure}

As a direct consequence of ``line interlocking", a line photon emitted by one ion can be absorbed by another ion of the same/different element with approximately the same energy. 
This section discusses the atomic processes following absorption of the lines in the Fe~XXV~K$\alpha$ complex by
various ions, while section \ref{Spectroscopic evidence} explores the changes in Fe XXV K$\alpha$ line intensities
due to the absorption.

The top panel of Figure \ref{fig:FeXXIII_FeXXIV_FeXXV_FeXXVI} shows the energy levels for one-, two-, three-, 
and four-electron iron. The red dashed line marks the energy of the Fe XXV K$\alpha$ complex. The Figure shows  a
clear proximity of selected Fe$^{22+}$ and Fe$^{23+}$ energy levels to the Fe XXV K$\alpha$ energies.
These energy levels lie well above the ionization limits of three-, and four-electron iron, hence are autoionizing levels. 
An enlarged version near the Fe~XXV K$\alpha$ energies is shown in the bottom panel 
of Figure \ref{fig:FeXXIII_FeXXIV_FeXXV_FeXXVI}. An additional column showing the energy levels in Cr$^{22+}$ 
is included in the enlarged version as the x line-center optical depth has contributions from  a Cr XXIII transition too (see section \ref{Line interlocking}).
The probability of a Fe XXV K$\alpha$ line photon being absorbed by other ions depends on two factors.
First, whether the energy levels of the absorbing ion are similar to the emitted line photon to allow line interlocking. The bottom panel of Figure \ref{fig:FeXXIII_FeXXIV_FeXXV_FeXXVI} gives a visual illustration of the proximity in energy levels in different ions. 
Second, if the optical depth for that transition is large enough for the absorption to happen, where optical depth is given by the equation:
\begin{equation}
\label{eqn:LineAbsorptionCrossSection}
 {\tau _\nu(x) } = 2.24484 \times 10^{-14} A_{u,l} \lambda_{\mu m}^{3}\frac{{g_u}}{{g_{l}}} \frac{\varphi _\nu(x) }{u_{\rm Dop}} \rm{N_{ion}}
\end{equation}

Once absorbed, the excited ion may or may not undergo autoionization.
The probability of autoionization depends on the following ratio: $p_{a}=\frac{A_a}  {{(A_a+ A_{u,l})}}$, 
where A$_a$ is the autoionization rate of the upper level of the excited ion, and  A$_{u,l}$
is the downward radiative transition probability. For light elements like helium, $p_{a}$ is usually close to 1. In heavier elements like iron, $p_{a}$ can be as small as $\sim 10^{-5} $ for some levels. Table \ref{t:E_Fe XXIV} gives the list of A$_{u,l}$, and A$_{a}$ for the Fe$^{23+}$ , Cr$^{22+}$,  Fe$^{22+}$ energy levels in close proximity to the Fe~XXV~K$\alpha$ energies, and the list of  A$_{u,l}$ for Fe$^{24+}$ generating the Fe XXV K$\alpha$ complex.

The total x line-center optical depth becomes $\geq$ 1 only for hydrogen column densities - N$_{\rm H}$ $\geq$ 10$^{24}$ cm$^{-2}$.
The transition wavelengths in the proximity of x ($\lambda$= 1.8554\AA) are:
two Fe XXIV  ($\lambda$= 1.8563\AA, 1.8571\AA), and one Cr XXIII ($\lambda$= 1.8558\AA) satellite lines.
Among these, one of the Fe XXIV satellite ($\lambda$= 1.8563\AA) 
has a negligible contribution in the absorption of x due to its small A$_{u,l}$ (see table \ref{t:E_Fe XXIV}). 
The transitions dominating absorption of x are therefore Fe XXIV ($\lambda$= 1.8571\AA), 
and Cr XXIII ($\lambda$= 1.8558\AA). The former has $p_{a}=0.29$, while the latter has $p_{a}\sim 10^{-2}$ 
for one scattering. Therefore, in the instances of a single absorption by Fe$^{23+}$
corresponding to $\lambda$= 1.8571\AA, approximately one-third of  Fe$^{23+}$  will autoionize
destroying the absorbed x photons, causing a deficit in the x line intensity. If absorbed by Cr$^{22+}$ , 
the ion will most likely de-excite by the emission of a line photon of the same wavelength having no effect on the x line intensity.

 In contrast, the line-center optical depth in y ($\lambda$=1.8595\AA) has a small contribution
 from the Fe~XXIV satellite at the wavelength 1.8610\AA~with $p_{a}\sim 10^{-5}$. Even though a small fraction of y is absorbed by Fe$^{23+}$, the
 autoionization probability is negligible.
Up to very high column densities (N$_{H}$ $\leq$ 10$^{25}$ cm$^{-2}$) the  line-center optical depth in z remains $<$ 1. Thus, the absorption of z photons ($\lambda$=1.8682\AA) does not become important despite its proximity with the Fe XXIII satellite line at $\lambda$=1.8704\AA. Beyond this column density, however, some z photons will be absorbed by Fe$^{22+}$, with a
very small probability of autoionization and loss of z photon ($p_{a}\sim 10^{-2}$).



\begin{figure*}
\gridline{\fig{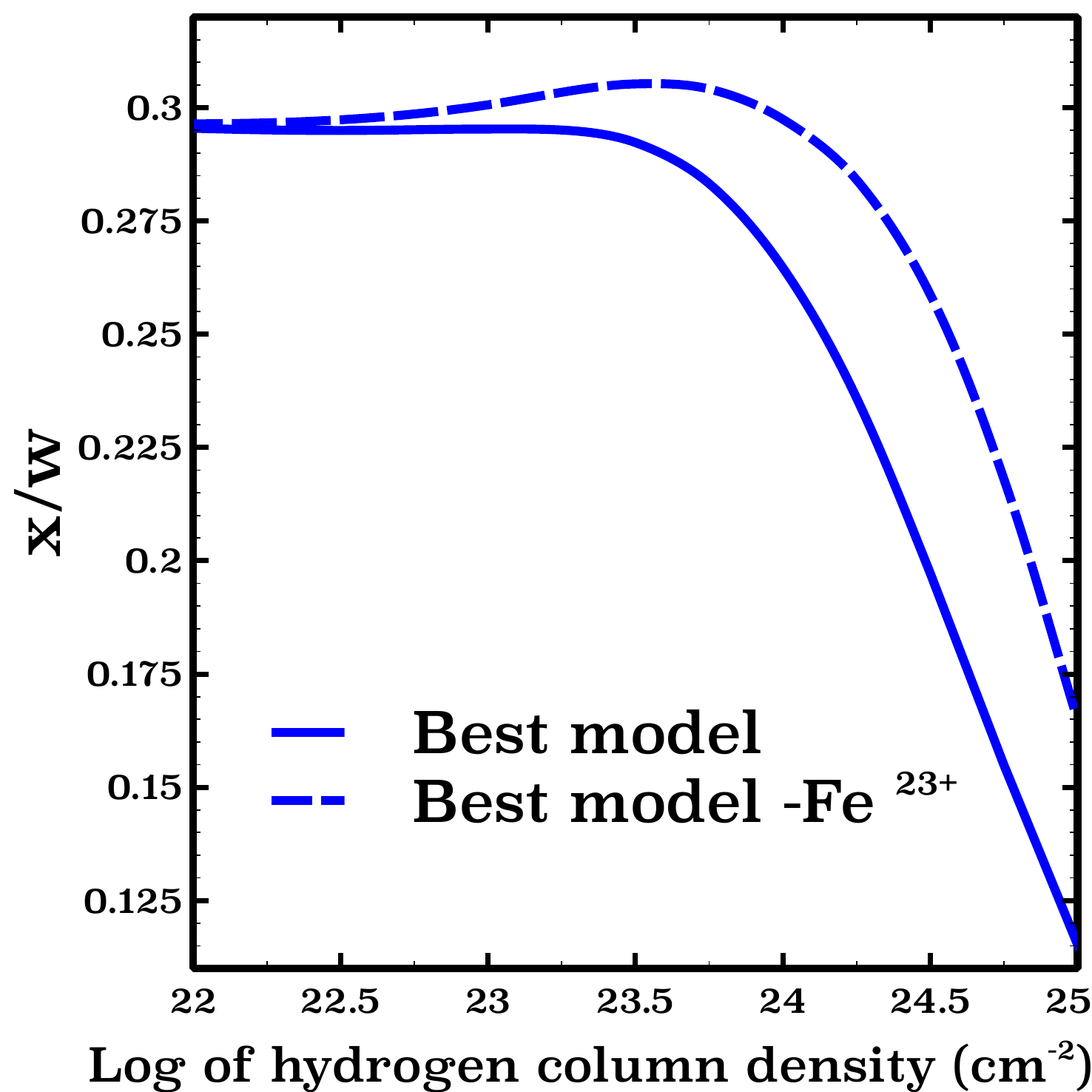}{0.32\textwidth}{(a)}
          \fig{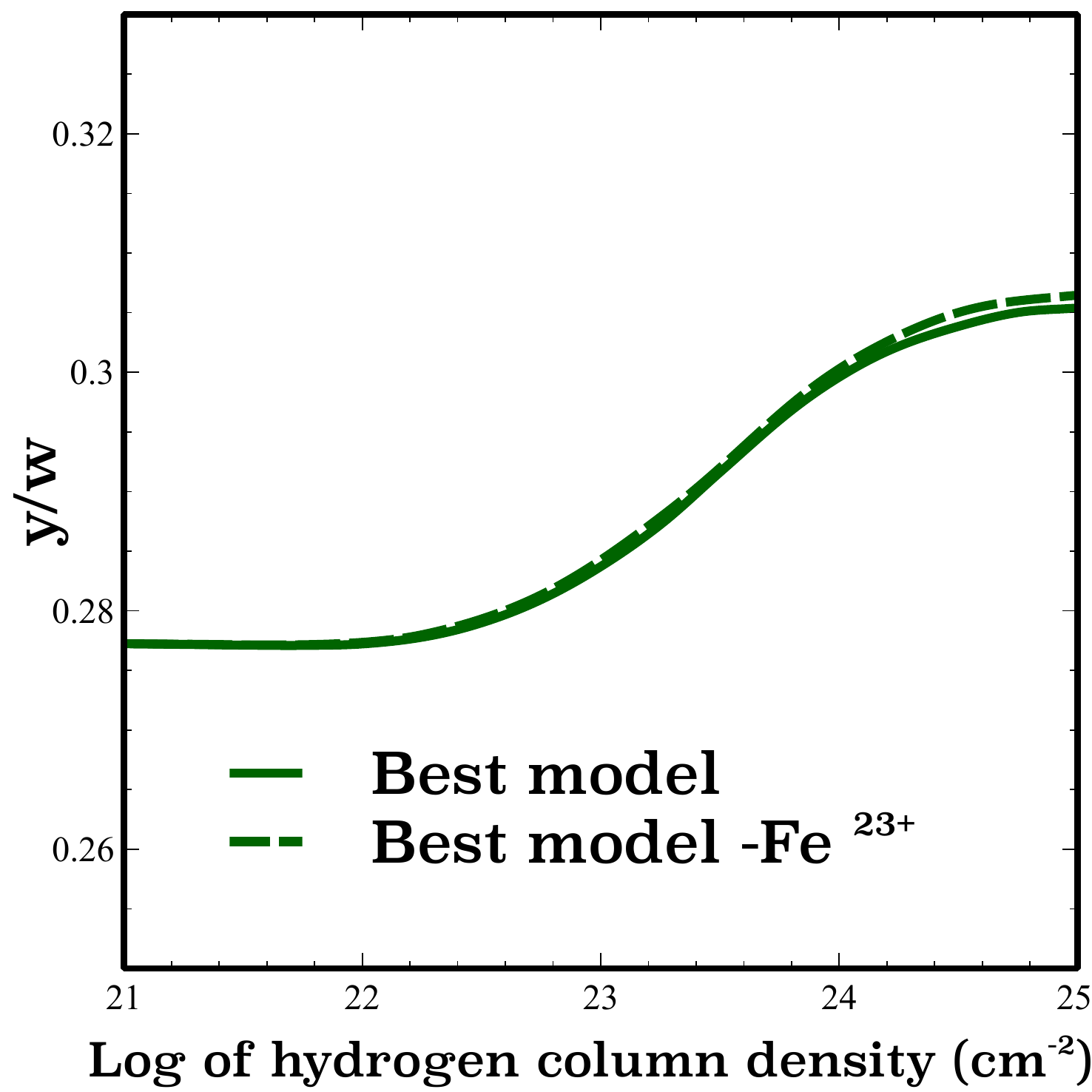}{0.32\textwidth}{(b)}
          \fig{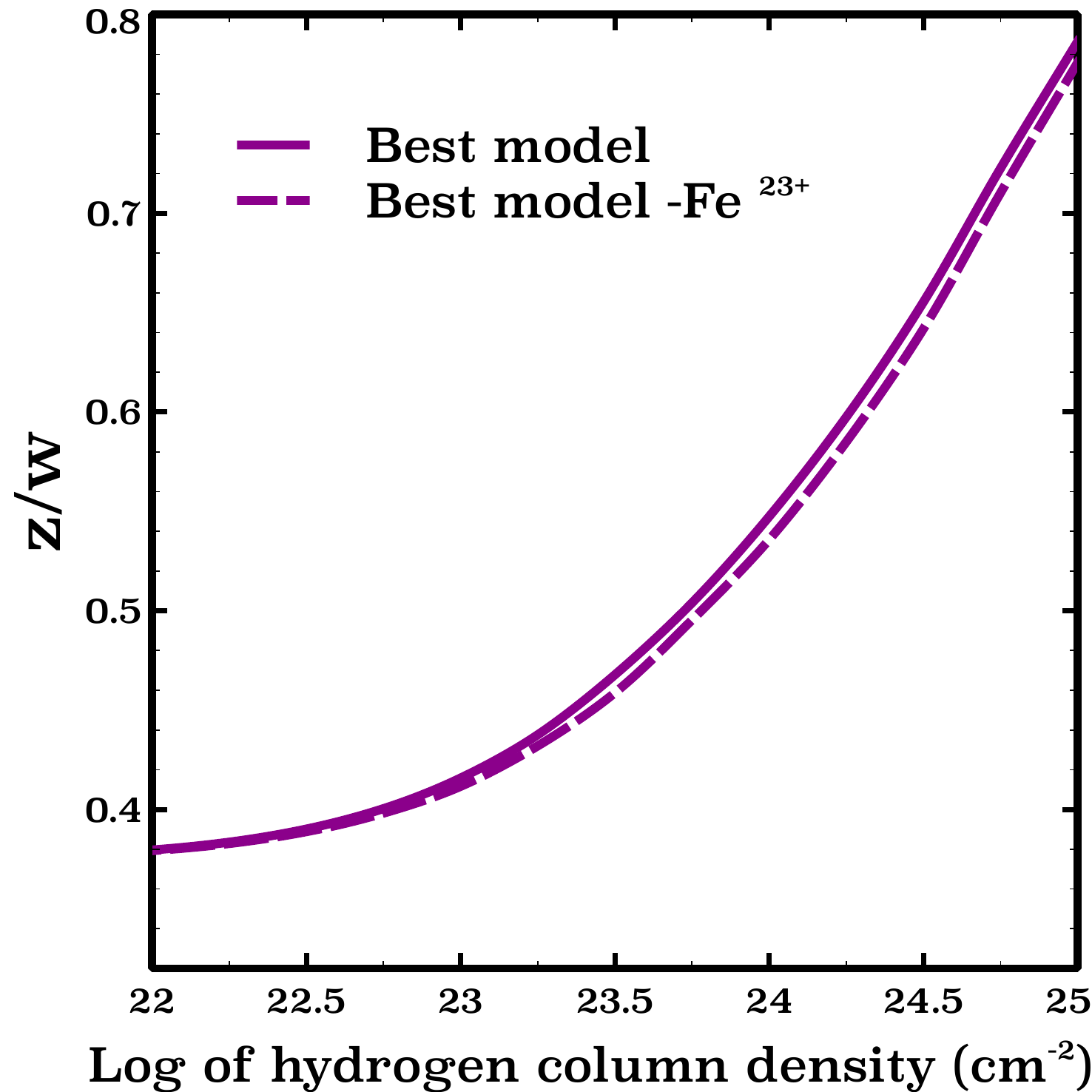}{0.32\textwidth}{(c)}
          }
\gridline{\fig{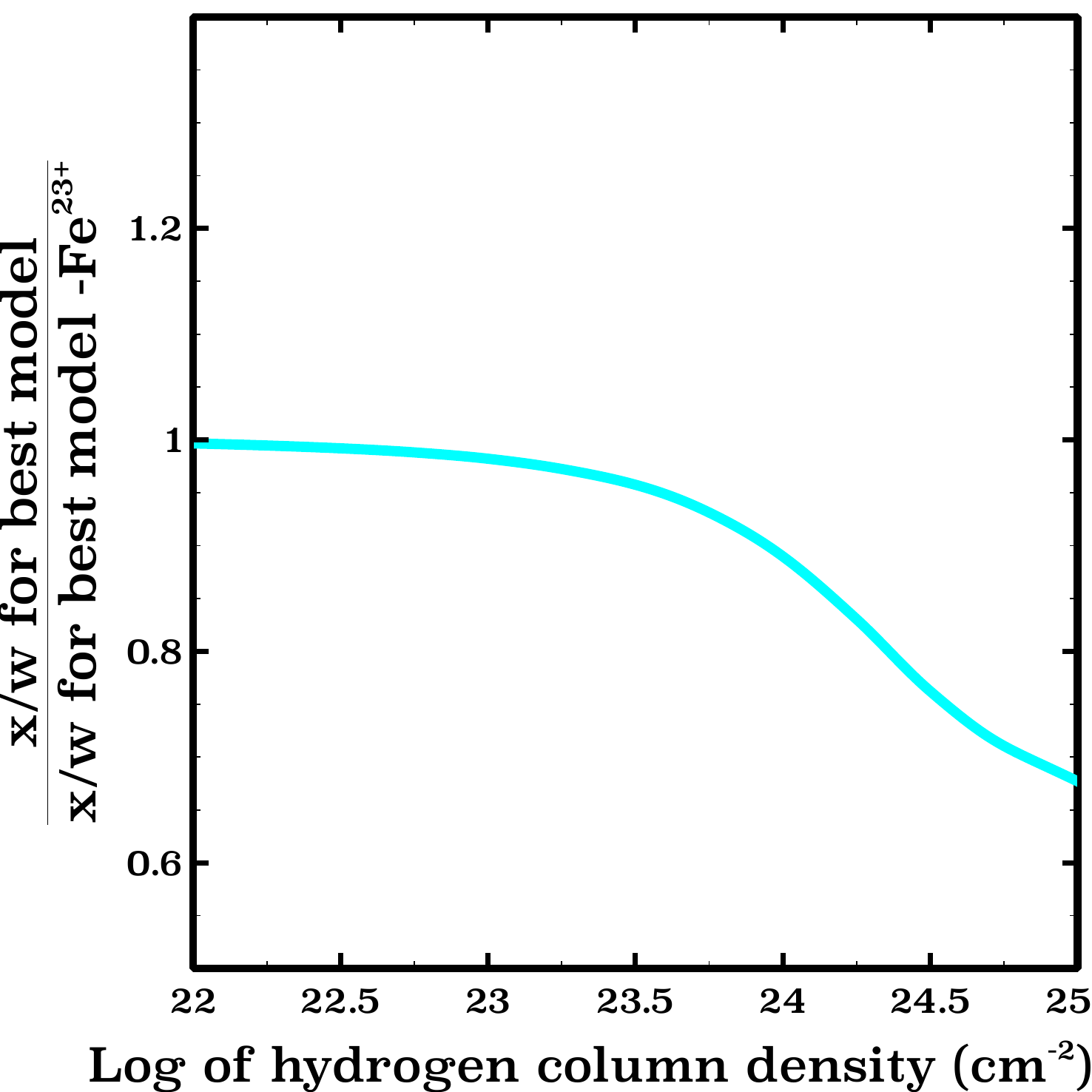}{0.32\textwidth}{(d)}
          \fig{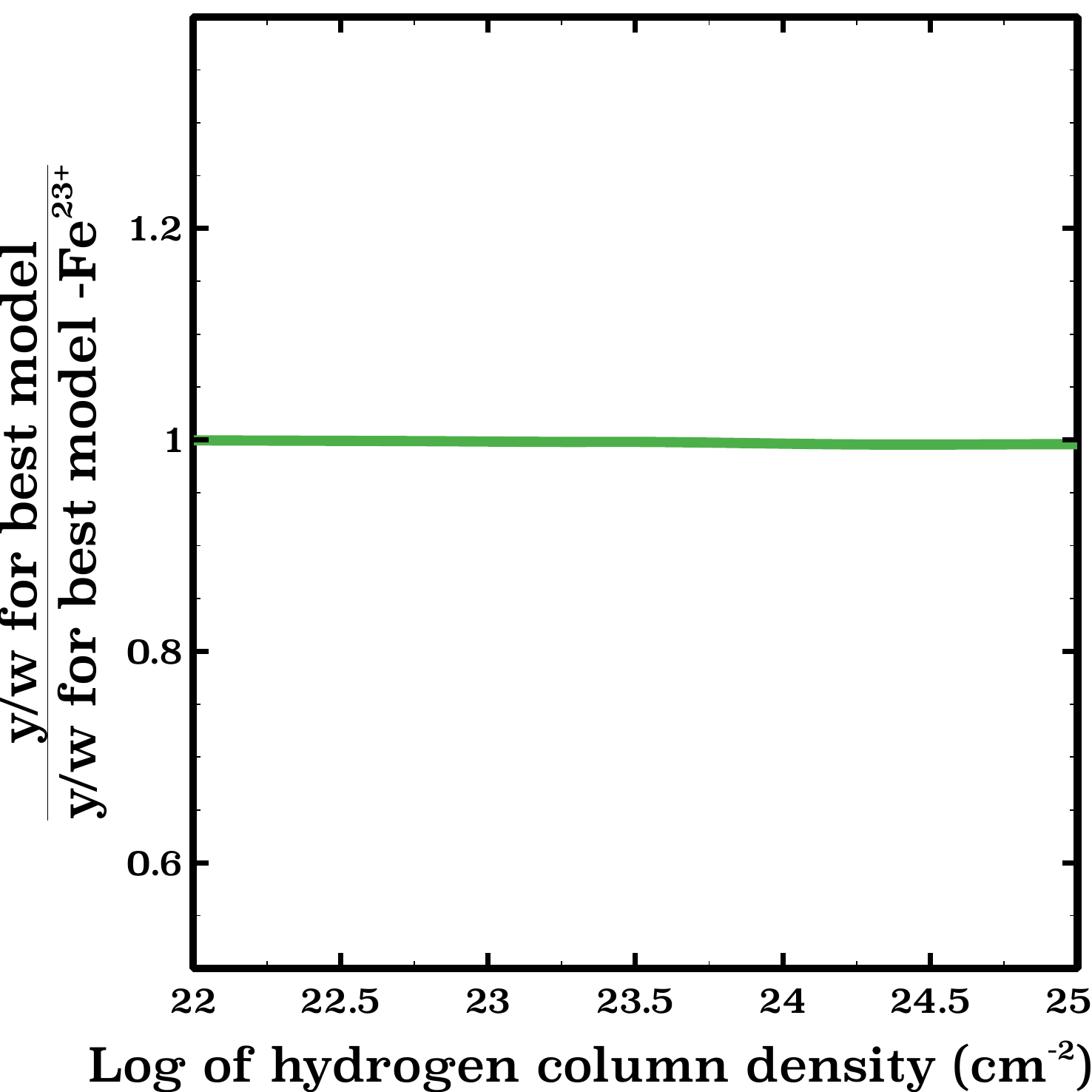}{0.32\textwidth}{(e)}
          \fig{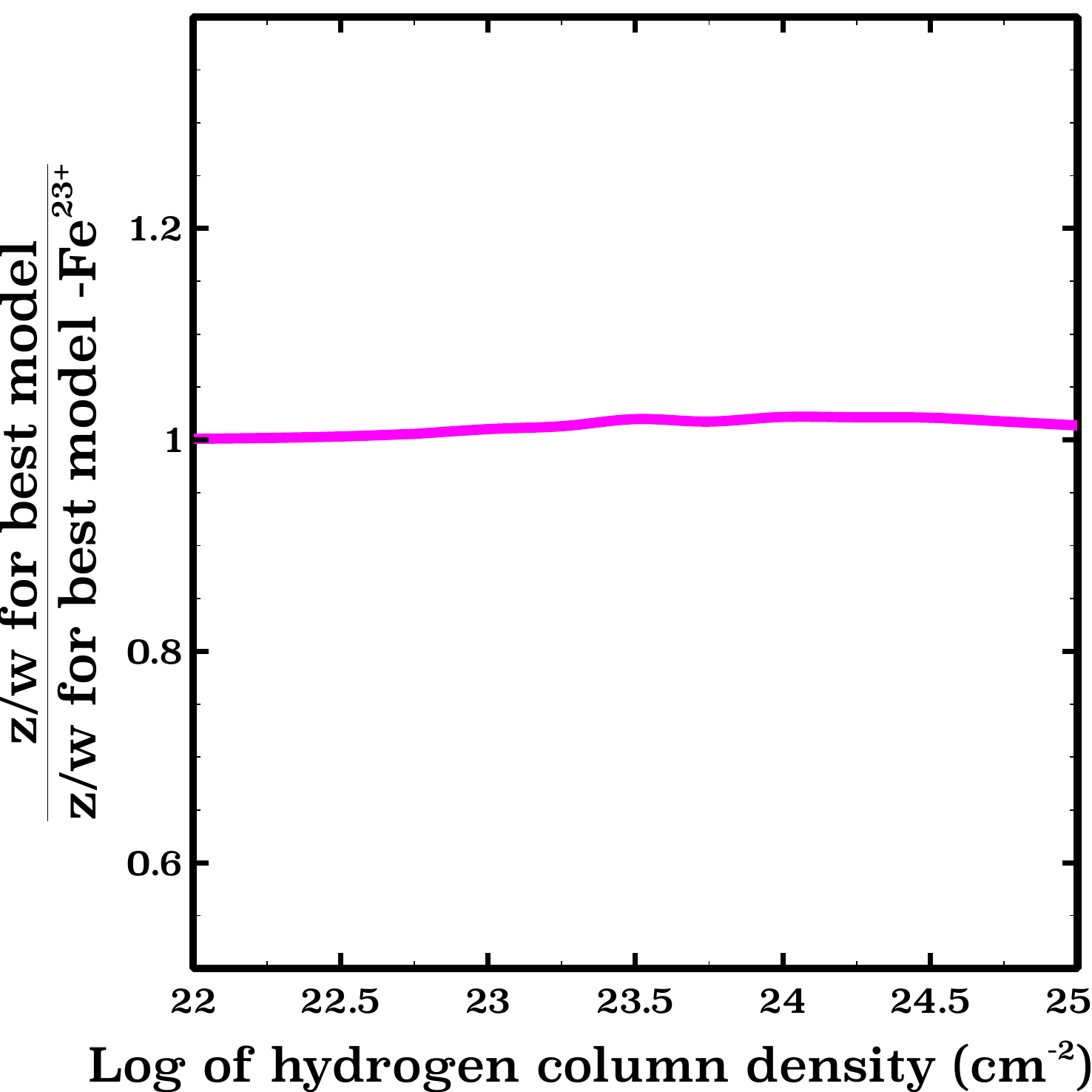}{0.32\textwidth}{(e)}
          }
\caption{Top: Variation of x/w, y/w, and z/w with hydrogen column density for our best model, and a model excluding Fe$^{23+}$.
 Bottom: Variation of the ratios of our best model to the model without Fe$^{23+}$ for x/w, y/w, and z/w  with hydrogen column density.}

\label{f:withwithoutlaser}
\end{figure*}


The loss of line photons following autoionization is referred to as Resonance Auger Destruction (RAD) in \citet{2005AIPC..774...99L}.
This not only leads to suppression of selective line intensities in the high column density (see the next section) 
but also causes the absorbing ion concentration to decrease due to autoionization. 
The higher the p$_{a}$, the higher the number of RAD and the more pronounced this effect will be. 
That is why the x line-center optical depth shows a clear slower-than-linear growth in the high column density
range (N$_{H}$ $\geq$ 10$^{24}$ cm$^{-2}$) in the right panel of Figure \ref{f:Lineinterlockig},
while y is only slightly deviates. In the very high column density (N$_{H}$ $\geq$ 10$^{25}$ cm$^{-2}$), 
line-center optical depth in z shows a slight slower-than-linear growth due to absorption by Fe$^{22+}$, following RAD.


\section{Spectroscopic Evidence of RAD}\label{Spectroscopic evidence}

The continuous increase in the line intensities with column density makes it difficult to detect 
where the redistribution in photon energy 
is happening via absorption/re-emission (see Figure \ref{fig:absolute}). Such effects can be best demonstrated with a line-ratio
diagnostic instead.

We plot the variation of x, y, z line intensities relative to w  versus hydrogen column density  in Figure \ref{f:withwithoutlaser}.
As the total-line-center optical depth in w comes entirely from its single-line optical depth, 
absorption by other ions does not have any effect on its line intensity. 
The amount of change in the line ratios of x, y, and z relative 
to w will, therefore, only be subject to the change in the numerator intensities due to absorption by ions.

In the low-column-density limit, all the lines in the Fe XXV K$\alpha$ complex are optically thin.
Line photons emitted in the cloud escape. Without absorption by ions, the line intensities
increase with column density. The line ratios, therefore, remain unchanged in the low-column-density limit. 
In the high-column-density limit, two atomic processes simultaneously change the line
intensities of the Fe XXV K$\alpha$ complex: absorption of line photons following line interlocking, 
and  the transfer from Case A to B. The former is studied in this paper, while the latter is the subject of the accompanying paper (paper II).
Besides, in the very high-column-density limit (N$\rm_{H}$ $\geq$ 10$^{23}$ cm$^{-2}$), ESE also starts to become important.


Section \ref{Line interlocking} discussed the contribution of absorption by Fe$^{23+}$ in deciding the 
line-center optical depth in x and y. CLOUDY normally determines the population of various ion stages self consistently. To test the role of  Fe$^{23+}$, we artificially set the abundance of Fe$^{23+}$ to a very small value and compared it with an all-ion-inclusive model.
The top panel of Figure \ref{f:withwithoutlaser}  compares the x/w, y/w, and z/w ratios with and without Fe$^{23+}$ . 

The top-left panel of the Figure compares the x/w ratios with and without Fe$^{23+}$. The change in the x  line intensity for our best model (with Fe$^{23+}$)
 comes from the Case A to B transition, line interlocking with Fe$^{23+}$, and ESE (see section \ref{ESE}), and other factors mentioned in the appendix. The Case A to B transition causes all 2 $\rightarrow$ 1 line intensities,  and therefore the line ratios, to increase (see section 6.2 in paper II). We find the increase in x/w due to Case A to B transfer to be small. The decrease in the x/w ratio with column density in the  Fe$^{23+}$ inclusive model is partly due to 
interactions with Fe$^{23+}$. As w is not affected by Fe$^{23+}$, the decrease in the x/w line ratio in 
the presence of Fe$^{23+}$ only reflects the deficit in x line photons at the higher column densities. According to our calculations, $\sim$ 32\% percent of x photons are lost due to absorption by Fe$^{23+}$ at the hydrogen column density 10$^{25}$ cm$^{-2}$.

Analytically, the fraction of photons destroyed due to RAD depends on the downward radiative transition 
probability (A$_{u,l}$), and autoionization rate of the upper level (A$_{a}$). When the x photon
is absorbed by Fe$^{23+}$, corresponding to the wavelength 1.8571\AA, the fraction of x photons destroyed in 
one scattering is: $p_{a}= \frac{A_a}{(A_a+ A_{u,l})} = 0.29$ as discussed in the previous section. The optical depth contribution at the line center of x due to line interlocking with the  $\lambda$= 1.8571\AA$\hspace{0.2mm}$
  transition is $\sim$ 1 at the hydrogen column density of 10$^{25}$ cm$^{-2}$. This implies that $\sim$ 29\% of the x photons are destroyed at N$_{\rm H}$ = 10$^{25}$ cm$^{-2}$. The analytical theory of RAD is, therefore, 
consistent with our calculation, which predicts a 32\% deficit in x line intensity at the same hydrogen column density.

For the sake of simplicity in the demonstration, we compare  the analytical and calculated values for the decrease in x line intensity due to RAD at the hydrogen column density 10$^{25}$ cm$^{-2}$. RAD in x however, begins at a much lower hydrogen column density (N$_{\rm H}$ = 10$^{23}$ cm$^{-2}$ ). The percentage of x photons destroyed at the hydrogen column densities 10$^{23}$ cm$^{-2}$, 10$^{24}$ cm$^{-2}$, 10$^{25}$ cm$^{-2}$  are $\sim$ 2-3\%, $\sim$ 12\%, and $\sim$ 32\%, respectively. The bottom panel of Figure \ref{f:withwithoutlaser} shows the variation of the ratios of x/w, y/w, and z/w for our best model to the model without Fe$^{23+}$ with hydrogen column density.

The upper middle panel of Figure \ref{f:withwithoutlaser} compares the y/w ratio for our best model and
a model excluding Fe$^{23+}$. Unlike x, the presence of Fe$^{23+}$ slightly alters y line intensity ($\leq$ 2\% ). 
This is expected, as the line-center optical depth in y has much smaller contribution 
from  Fe XXIV ($\lambda$= 1.8610\AA), with $p_{a}\sim 10^{-2}$.
The slight increase in y/w in the high column density happens due to Case A to B transfer and will be addressed in paper II.

The right panel of Figure \ref{f:withwithoutlaser} compares the z/w ratio with and without Fe$^{23+}$. As the presence of Fe$^{23+}$ 
does not make z
optically thick even at our highest column density limit (N$_{\rm H}$=10$^{25}$ cm$^{-2}$) , ideally z line intensity should 
remain the same for the with or without Fe$^{23+}$ models.
However, both  plots in the right panel show a small increase in the z/w ratio ($\sim$ 2 \%) in the presence of Fe$^{23+}$. This behavior is 
 opposite to that of x/w indicating that the number of x photons lost is converted to z photons causing z/w to 
increase in the with Fe$^{23+}$ model. This is Case A to B behavior discussed in detail in paper II, and points towards the conversion of a tiny fraction of x photons into z photons.

\section{Electron scattering escape}\label{ESE}
\subsection{General Formalism}\label{general formalism}
The mean number of scatterings (N) experienced by a line photon before escaping an optically thick cloud is related to its
line-center optical depth ($\tau$) with the following equation \citep{1979ApJ...229..274F}:

\begin{equation}
\label{eqn:N}
\rm{N} =
\begin{cases}
1.11\tau ^{0.826} & \text{if $\tau < 1$} \\
\frac{{{ 1.11\tau ^{1.071}}}}{{1+{\rm{({log\tau/5.5}})^5}}} & \text{if $\tau \geq 1$}
\end{cases}
\end{equation}
Notice that the number of scatterings is nearly linear in $\tau$ and is not proportional to
$\tau^2$, as would occur if it did a random walk in space.

Cloudy uses the formalism developed by \citet{1980ApJ...236..609H} 
in treating energy loss during resonance scattering.  
This treatment does apply to Fe K$\alpha$ since the range
of damping constants they consider does extend up to the very large values
($a \sim 0.045$ for w) found in allowed X-ray transitions.  
This theory predicts a modest increase in path length, which
is taken into account.

The presence of  fast thermal electrons in the cloud may lead to
one-scattering-escape in some fraction of the line photons. 
This happens when line photons receive a large Doppler shift from their line-center upon scattering off  high-speed electrons.
The probability of scattering off electrons is given by the following:
\begin{equation}
\label{eqn:ESCprob}
\rm {P}_{\rm broad} = \frac{{{ \kappa_{e}}
}}{{\kappa_{tot}}}
\end{equation}
\noindent 
where  $\kappa_{e}$ is the electron scattering opacity, and  $\kappa_{tot}$ is the total line-center opacity at that wavelength. $\kappa_{tot}$ includes total line opacity ($\kappa_{line}$) and continuous opacity ($\kappa_{con}$). The latter includes electron scattering ($\kappa_{e}$), photoelectric absorption, and dust opacity (if dust is present)\footnote[1]{ \scriptsize{\footnoterule{  We evaluate equation 3, assuming that the line opacity does not depend on optical depth.  This is equivalent to assuming that the opacity does not depend on frequency.  This is true for electron scattering since the Klein-Nishina cross-section varies very slowly with energy.  The K$\alpha$ line opacity has an energy dependence given by the Voigt function.  Lines that scatter in the core of the profile undergo complete redistribution where the absorbed and emitted photon energy is uncorrelated.  This leads to a frequency-independent source function, one that is constant across the Doppler width.  Thus, to a fair approximation, we can evaluate the line opacity and assume that the ratio given in equation 3 does not depend on optical depth. We know of no numerical calculations of transport of strongly damped lines with background opacity, other than the \citet{1980ApJ...236..609H}  work, which considers only larger optical depths.  Such calculations should be done in support of future microcalorimeters missions.}}}.

The fraction of photons escaping the cloud after N scattering is :
\begin{equation}
\label{eqn:frac}
\rm {f_{ese}} = \rm 1- (1-P_{\rm broad})^{N}.
\end{equation}
We name this process of photon escape following a single event of scattering off electron- ``Electron  Scattering Escape (ESE)". 

At the Perseus core temperature, the velocity of an electron is mildly relativistic ($\sim$ 37200 km/s), about 15\% of the speed of light.
The ratio of electron to iron thermal widths is $\sqrt{m_{\rm Fe} / m_{\rm e}}$ $\sim$ 300. When the photons scatter of such high-speed electrons, they are Doppler shifted far away from the line center and escape the cloud. This leads to the broadening of Gaussian line profiles with a super-broad-base under Fe XXV K$\alpha$ of $\sim$ 1 keV width. 
This is a line broadening process that can be observed only in the  high-column-density 
limit with an optically thick cloud. Therefore, Perseus is not subject to line broadening due to ESE. But optically thick environments such as accretion disks will show such line broadening. 

\begin{figure*}
\gridline{\fig{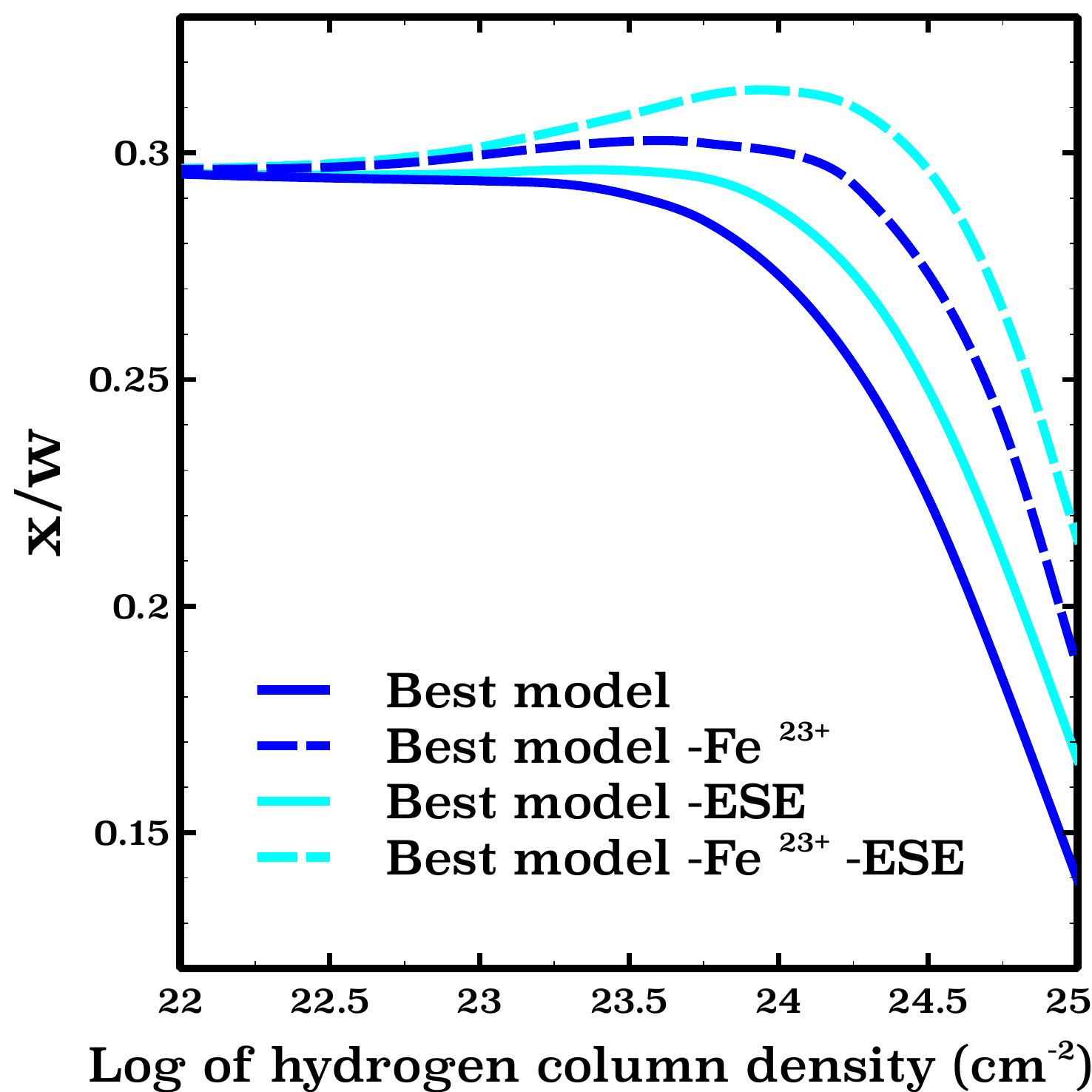}{0.32\textwidth}{(a)}
          \fig{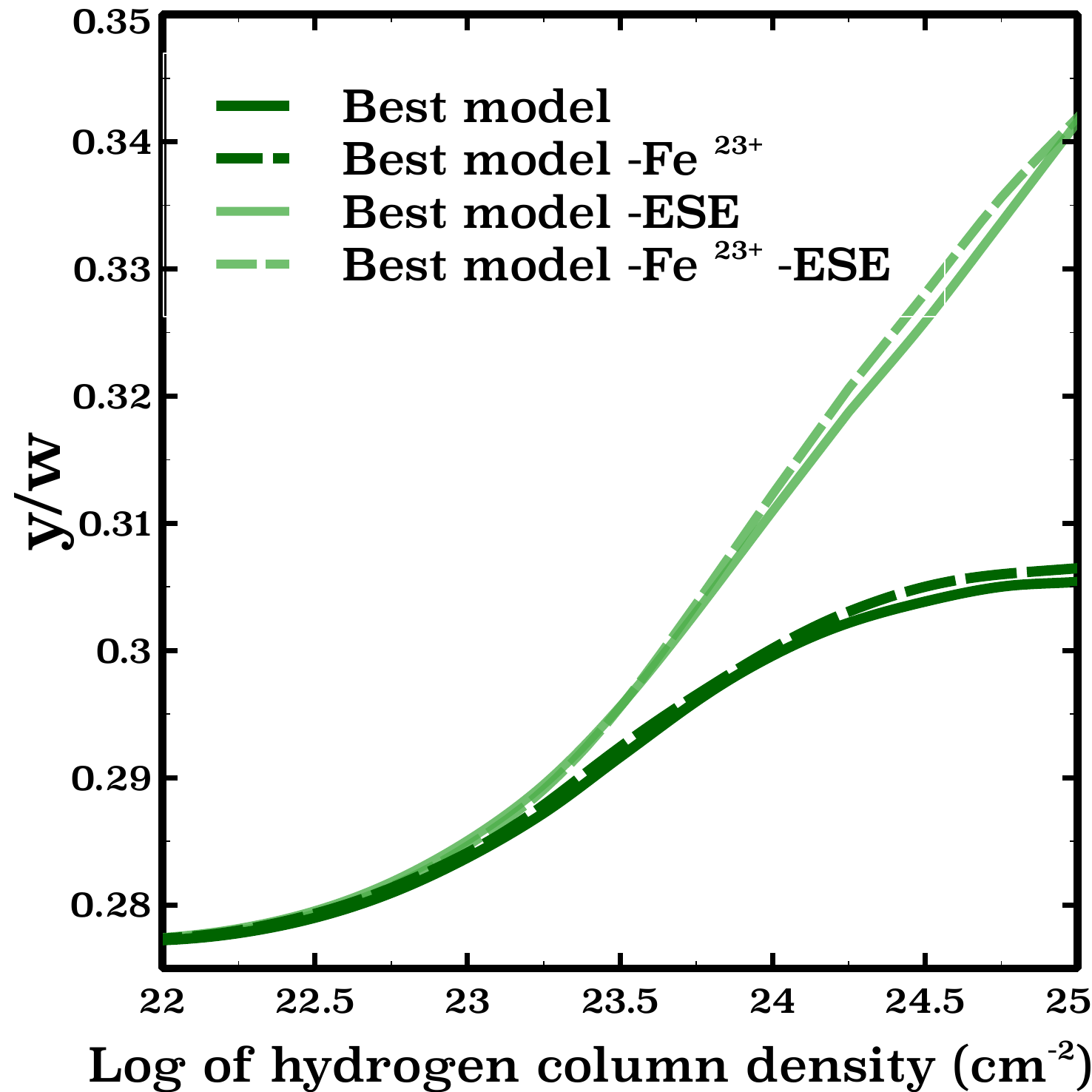}{0.32\textwidth}{(b)}
          \fig{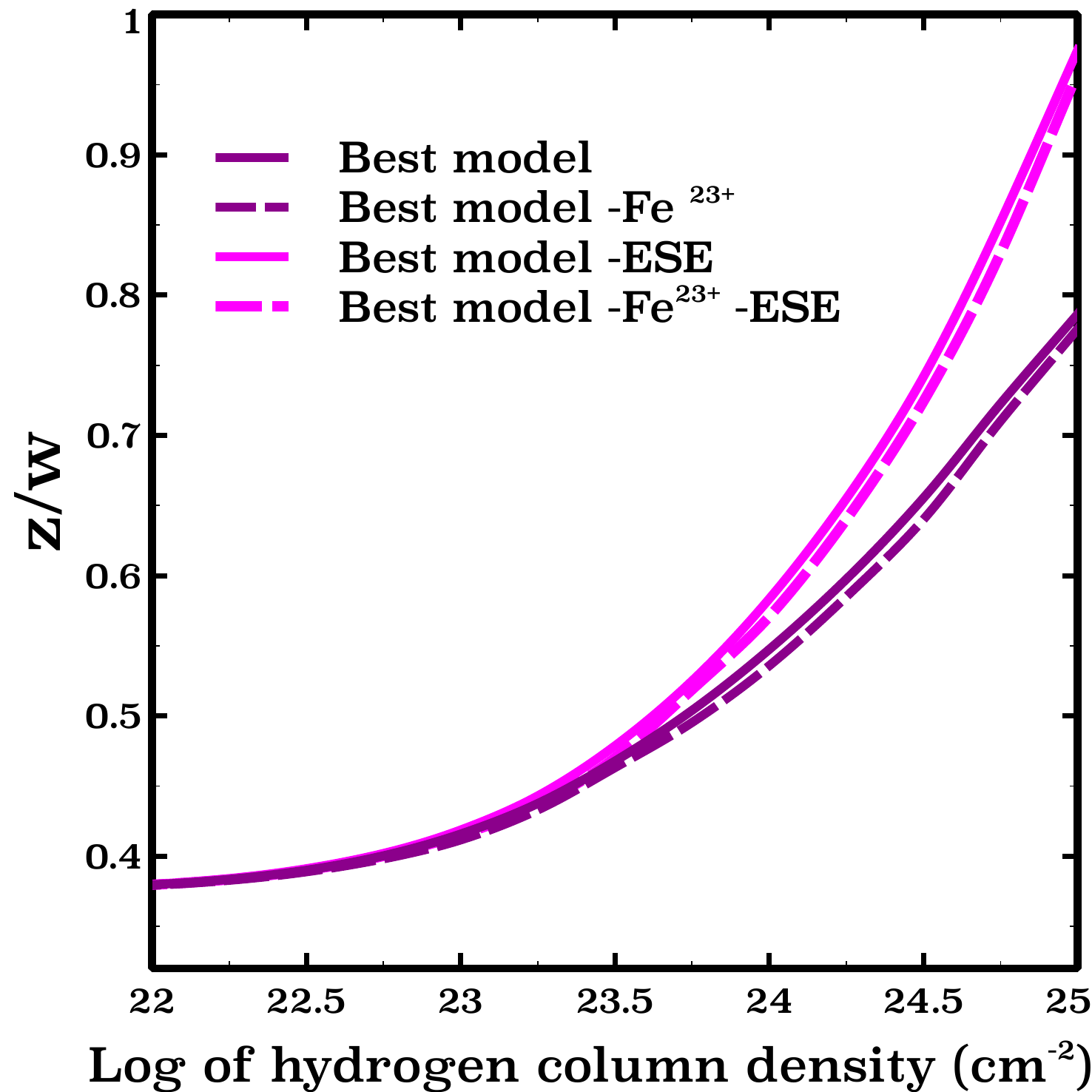}{0.32\textwidth}{(c)}
          }
\caption{Left, middle and right panels  show the variation of x/w, y/w, and z/w  with respect to the hydrogen column density for our best model and a Fe$^{23+}$ exclusive model, with and without ESE.
The best model consists of a narrow line component and a broad line component coming from ESE.
\label{f:EES}}
\end{figure*}

ESE depends on the two following factors:  the probability of being scattered by electrons (P$_{\rm broad}$), and the mean number of scatterings (N) experienced by the line photons (see equation \ref{eqn:frac}). 
The single-line opacities ($\kappa_{line}$) in x, y, and z are smaller than that of w.   
Their P$_{\rm broad}$ values are larger, which makes them more likely to suffer electron scattering.
But w  has a much ($\sim 100$ or higher) larger mean number of scatterings N, which tends to increase
the possibility of ESE, but its much larger $\kappa_{line}$ makes P$_{\rm broad}$ smaller, which tends to decrease ESE.
This complex interplay between N and P$_{\rm broad}$ causes  changes in the line profiles in x, y, z, and w. However, this does not change
 the line intensities in x, y, z, or w.  The Gaussian line profiles get broader and flatter due to ESE, keeping the area under the Gaussian and the intensity constant. We show the ratio of the total line intensities under the Gaussians in Figure \ref{f:EES}. The line intensities (and the line ratios), although independent of ESE in x, y, z, and, w, depend on the following factors:

First, ESE in other line photons can cause the line intensities in x, y, z, and w to change. 
Figure \ref{f:EES} compares the line ratios with respect to w with    
and without ESE for the following cases:
a) in an all-ion-inclusive model /our best model, and b) for a model excluding Fe$^{23+}$. 
In the optically thick (Case B) limit, 
higher $n$ Lyman lines are converted to Balmer plus Ly-$\alpha$ photons, causing an overall 
increase in the Fe XXV K$\alpha$ complex line intensities (paper II). 
Higher-$n$ Lyman lines have smaller $\kappa_{line}$, 
and so have a higher probability of being scattered by electrons according to equation \ref{eqn:ESCprob}. 
A fraction of the higher-$n$ Lyman lines will, therefore, escape the cloud through ESE before being converted to Ly-$\alpha$. This will result in a net decrease
in the Fe XXV K$\alpha$ complex line intensities, which will weaken the weak $n=2 \rightarrow 1$ transitions (x, y, z) more than w, the strongest. This explains the overall decrease in x/w, y/w, and z/w line ratios in Figure \ref{f:EES} in the presence of ESE.

Second, the presence of ESE can slightly increase the line intensity in x. The x photons that escape the cloud following electron scattering will not be subject to absorption by Fe$^{23+}$. Thus ESE in x photons reduces the effect of RAD on x line intensity. This implies that the intensity of x line photons in our best model including ESE will be slightly enhanced compared to the without ESE model.
The line intensity change in Fe XXV K$\alpha$ complex due to ESE is collectively determined by ESE of higher $n$ Lyman line photons and weakening of RAD effects in x photons. Figure \ref{f:EES} reflects all these contributions.

Note that our best model in the Figure calculates line intensities including the super-broad-base under the Gaussian. Each Gaussian in our best model consists of two components. A broad line component
generating from ESE, and a narrow line component for the photons not scattered away from the line-center. A high-resolution X-ray telescope will only detect the narrow line component. But a low-resolution telescope with a resolution equivalent to 1 keV will detect the narrow and broad line components together. The difference in telescope resolution will, therefore, detect different line fluxes in the Fe XXV K$\alpha$ complex. In the next subsection, we discuss the variation of line ratios with column densities for the narrow line component, as a high-resolution X-ray telescope will observe it.


\begin{figure*}
\gridline{\fig{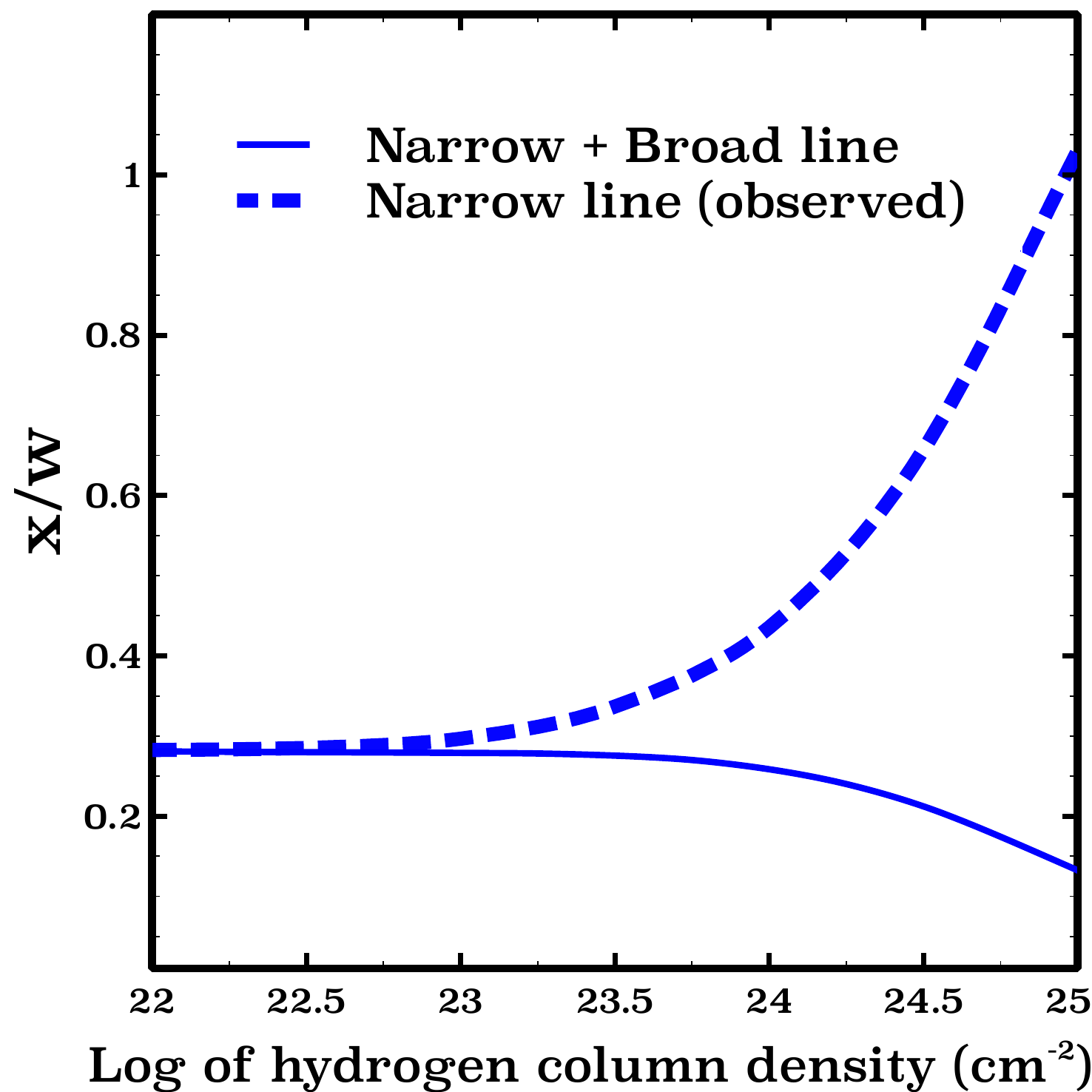}{0.321\textwidth}{(d)}
          \fig{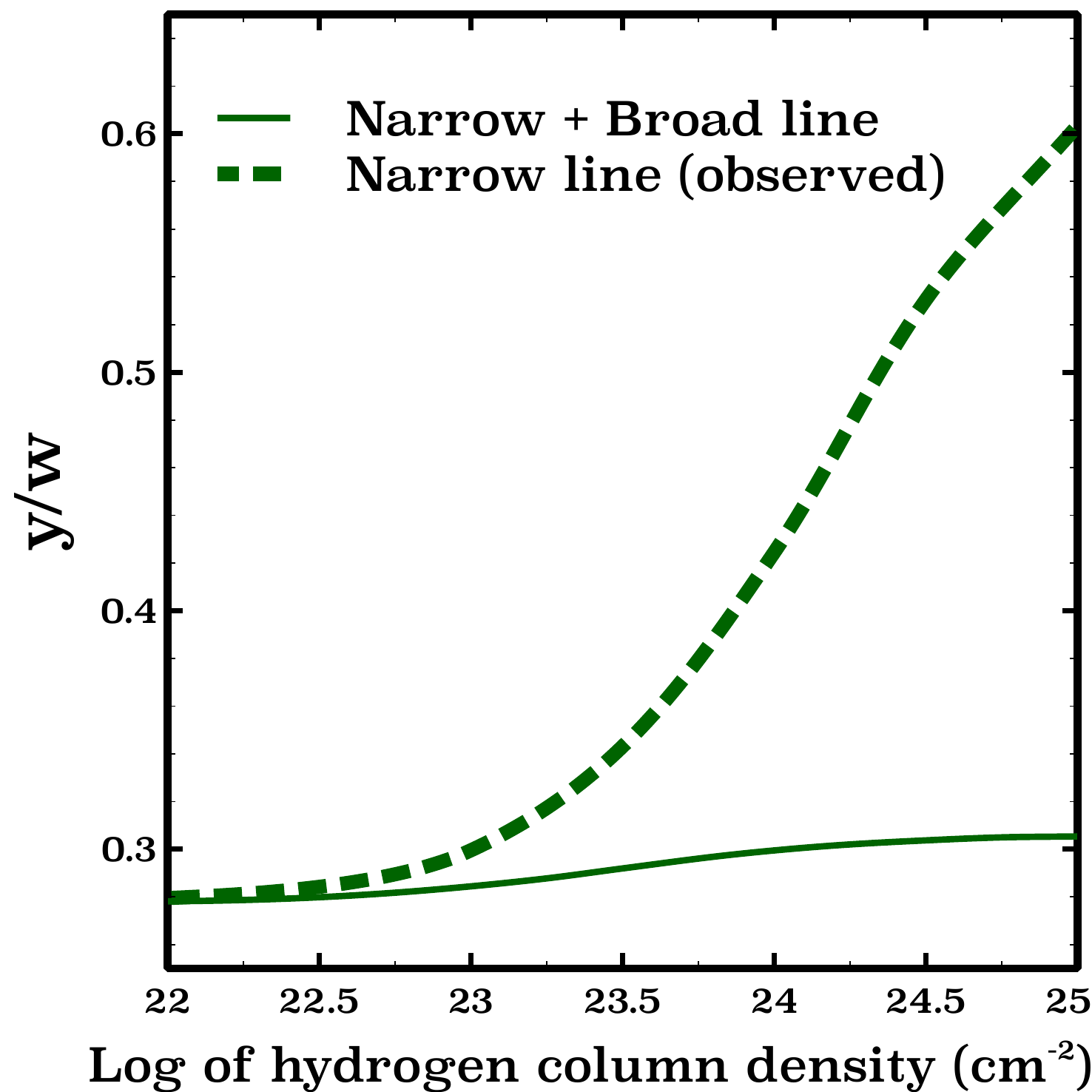}{0.321\textwidth}{(e)}
          \fig{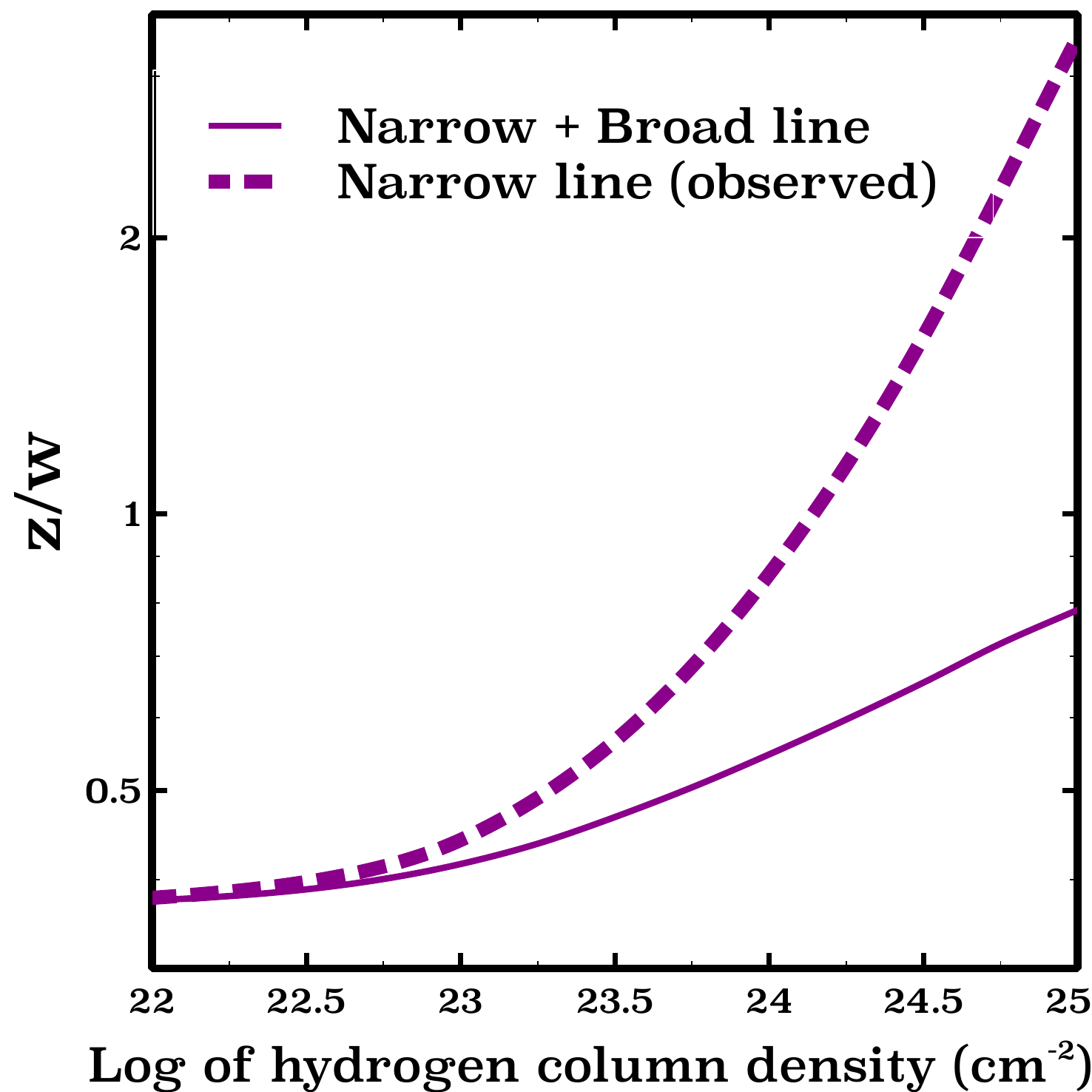}{0.321\textwidth}{(e)}
          }
\caption{Left, middle, and right panels show the variation of x/w, y/w, and z/w  with respect to the hydrogen column density for narrow + broad line, 
and narrow line components. The total effective intensity of the lines in the Fe XXV K$\alpha$ complex (and the line ratios) consists of the narrow + broad line component and
is the same as the best model shown in Figure \ref{f:EES}. 
The broad line component comes from ESE and has a spread of $\sim$ 1 keV. The narrow line component does not include the photons scattered by electrons and is the only line component 
detected by  microcalorimeters/ high-resolution X-ray telescopes.  The low-resolution telescopes will detect the narrow and broad line components together.
\label{f:EES2}}
\end{figure*}
\subsection{What will be observed?}

From the view of high-resolution X-ray spectroscopy, like Chandra HETG, or future observations by XRISM or Athena, photons that suffer ESE are lost from the Fe XXV K$\alpha$ complex. 
 Figure \ref{f:EES2}
shows the line ratios that will be observed in high-resolution. This is different from Figure \ref{f:EES}, which included the electron-scattered Fe XXV K$\alpha$ photons and the super-broad-base of the Gaussian in the calculation of line intensities (and line ratios). 

Let us analytically estimate the reduction in w line intensity in the observed high-resolution spectra due to the migration of w photons
away from its line-center, following ESE. The probability (P$_{\rm broad}$) that w suffers ESE and is lost from the observed spectra is given by equation \ref{eqn:ESCprob}. The probability that the w photons will be emitted as a sharp/narrow w line is given by : P$_{\rm narrow}$=1/N (N is calculated from equation \ref{eqn:N}). This narrow line intensity will be detected in the high-resolution spectra.  At N$_{\rm H}$=10$^{25}$ cm$^{-2}$, we find that : P$_{\rm broad}$ $\sim 5$  $\times$ P$_{\rm narrow}$. This implies that
the observed w line intensity at N$_{\rm H}$=10$^{25}$ cm$^{-2}$ should be 5 times smaller than its optically thin limit where the line photons are not subject to ESE. The increase in  x/w, y/w, and z/w in Figure \ref{f:EES2} partly comes from this reduction of w intensity in the observed high-resolution spectra. The net variation in the line ratios in the Figure is also partly due to a reduction in x, y,  and z intensity in the high-resolution observed spectra, and other factors contributing to the line intensity change explained in Section \ref{general formalism}. 





\section{Summary}\label{Summary}
This  paper  discusses the significance of three-electron iron in deciding the line intensities
of  the Fe XXV K$\alpha$ complex. We explore the prediction of \citet{2005AIPC..774...99L} on Resonance Auger Destruction (RAD)
with CLOUDY initially motivated by the Perseus cluster.  We extend our analysis to the wide range of column densities we encounter in astrophysics to which CLOUDY can be applied.
We summarize our results below:
\begin{itemize}
\item Whether a line photon escapes the cloud without scattering or is absorbed by ions of 
the same/different elements depends on its line-center optical depth. The total 
line-center optical depth corresponding to a transition can entirely come from that transition itself
or may have contributions from other transitions of the same/different element. A photon within a line scatters
off due to the total opacity,  with no knowledge of which transition or species created that opacity.
This leads to line interlocking with the possible loss of identity of the absorbed photon. Among the four members
in the Fe XXV K$\alpha$ complex, the x line-center optical depth has significant contributions from single 
line optical depths of selected Fe XXIV ($\lambda$= 1.8571\AA), and Cr XXIII ($\lambda$= 1.8558\AA) satellites
in very high column densities. y and z have contributions from  Fe XXIV($\lambda$= 1.8610\AA),
and  Fe XXIII ($\lambda$= 1.8704\AA) satellites, respectively. The line-center opacity in w is the same as its single-line opacity, and hence is unaffected. 
\end{itemize}

\begin{itemize}
\item Upon absorption by  Fe$^{23+}$, x photons excite  Fe$^{23+}$ to the 
autoionizing level  $ 1s.2s(\,\,^{3}S).2p \,\,^{2}P_{1/2}$ (corresponding to $\lambda$= 1.8571\AA). 
Theoretically, $\sim$ 29 \% of the x photon should be destroyed in one scattering/ for the optical depth of unity due to RAD.
Our calculations show very good agreement with theory. At the hydrogen column density of 10$^{25}$ cm$^{-2}$,
the optical depth contribution at the line center of x from  line interlocking with  $\lambda$= 1.8571\AA $\hspace{0.2 mm}$
transition is $\sim$ 1. At the same hydrogen column density, we see a  deficit of $\sim$ 32 \% 
 in the x line intensity in the presence of  Fe$^{23+}$ compared to the case with  Fe$^{23+}$  removed 
for our best model. 
The change in x line intensity due to RAD, however, begins at a much lower hydrogen column density (N$_{\rm H}$ = 10$^{23}$ cm$^{-2}$).  The percentage of x photons destroyed at the hydrogen column densities 10$^{23}$ cm$^{-2}$, 10$^{24}$ cm$^{-2}$, and 10$^{25}$ cm$^{-2}$  are $\sim$ 2-3\%, $\sim$ 12\%, and $\sim$ 32\%, respectively.

A small fraction of y photons excite Fe$^{23+}$ from the ground state to  $1s.2s(\,\,^{3}S).2p \,\,^{2}P_{3/2}$  ($\lambda$= 1.8610\AA),
but rarely get destroyed in RAD, because the autoionizing rate of the excited level is $\sim$ 10$^{-5}$ 
times smaller than the downward transition probability. 
z photons do not exhibit line interlocking with any Fe XXIV photons, 
and therefore the z line intensity is not directly affected by RAD.
\end{itemize}

\begin{itemize}
\item Line  photons can become  heavily Doppler-shifted  from  their  line-center  upon  
after scattering off fast thermal electrons that are present in the cloud. This leads to one-scattering-escape 
for a fraction of line photons, a process we call Electron Scattering Escape (ESE). 
Line broadening through this process becomes very important at high column densities. In the high-column-density limit, ESE can lead to a super-broad-base under Fe XXV k$\alpha$ of $\sim$ 1 keV width. The effective intensity of each line in the K$\alpha$ complex consists of a narrow line component and a broad line component coming from ESE. High-resolution  X-ray  telescopes like XRISM and Athena  will  only  detect  the  narrow line component, whereas  low-resolution telescopes  will  detect both  the narrow and broad line components. The line fluxes detected by high- and low-resolution telescopes will, therefore, be different.

\end{itemize}

\begin{itemize}
\item Line interlocking processes are very sensitive to line wavelengths. 
A change of $\sim$ 0.02\% in the wavelengths can significantly alter the line interlocking processes, 
as well as the nature of line intensities of Fe XXV K$\alpha$ complex.
In addition, the optical depth of a line with similar energy as that of Fe XXV K$\alpha$ complex determines the degree of interlocking,
 making  the accurate reporting of transition probabilities in the various atomic data sets extremely important.
We found a swap in the transition probabilities between the two Fe XXIV lines ($\lambda$= 1.8563\AA, $\lambda$= 1.8610\AA) in NIST. A swap in the transition probabilities will significantly change the x line-center optical depth, and change the hydrogen column density to 10$^{24}$ cm$^{-2}$ where we see a $\sim$50\% decrease in x intensity due to RAD. We verified this mistake with the
NIST team, and use CHIANTI version 9.0 instead for both wavelengths and  transition probabilities in our calculation.

\end{itemize}

\begin{itemize}
\item Although we  assume a static geometry in this paper, we point out that the physics of  RAD provides a diagnostic indicator of velocity gradients. 
The left panel of Figure \ref{f:withwithoutlaser} shows a deficit of $\sim$ 32\% in x line intensity in the presence of Fe$^{23+}$ due to RAD, as compared to the without Fe$^{23+}$ case. In our calculations for a static geometry, the artificial removal of Fe$^{23+}$  from our model negates the effects of RAD. But in a dynamic cloud, a velocity gradient between the Fe$^{23+}$ and Fe$^{24+}$ -emitting regions can also decrease/remove the RAD effects. For instance,  if the  Fe$^{23+}$ gas were moving by one or more thermal velocities relative to the Fe$^{24+}$ gas, there will be no overlap and no RAD. Such diagnostics will provide a possible indicator of the velocity gradients.
\end{itemize}

\begin{itemize}
\item The ratio of Fe$^{23+}$ to Fe$^{24+}$ abundance depends on the temperature of the plasma \citep{2013ApJ...768...82N}.
This ratio is higher in the low-temperature limit.  Therefore, the effects of line interlocking and RAD can be observed at much lower column densities in the systems cooler than Perseus. In such systems, there may be significant contributions from Fe$^{23+}$ and other satellite lines to what appears to be the x, y, and z.

\end{itemize}

\begin{itemize}
\item Our model assumes a symmetric emitting region, but the combination of asymmetry with the high w line optical depth will increase or decrease w relative to the other lines, as described by \citet{1987SvAL...13....3G}, and paper II.
\end{itemize}



\section*{Acknowledgement}

The comments of the referee were very helpful and added significantly to the presentation of our work.  His or her help is gratefully acknowledged. We thank Stefano Bianchi for his valuable comments. We acknowledge support by NSF (1816537, 1910687), NASA (17-ATP17-0141, 19-ATP19-0188), and STScI (HST-AR-15018).
MC also acknowledges support from STScI (HST-AR-14556.001-A).

\software{CLOUDY \citep{2017RMxAA..53..385F}}

\appendix
\section{The physics of the x transition}
\begin{figure}[h!]
\centering
\includegraphics[scale=0.5]{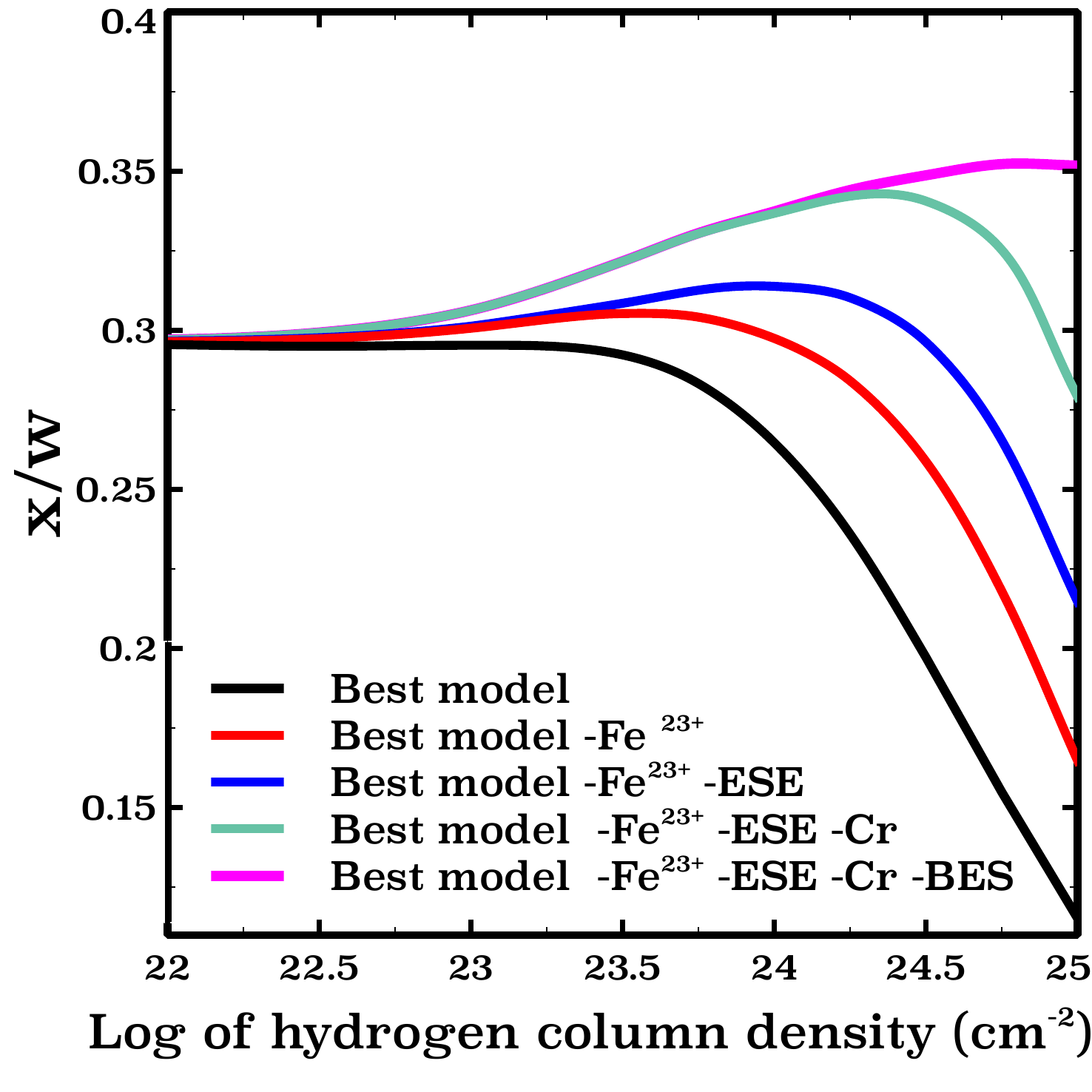}
\caption{The variation of x/w with hydrogen column density for a step by step removal of Fe$^{23+}$, electron scattering escape (ESE), Chromium, and broadband electron scattering (BES) from our best model.}
\label{fig:app}
\end{figure}

In the body of the paper, we discussed the effects of Fe$^{23+}$ and ESE in deciding the x/w ratio. However, two other factors change the line intensity in x — line interlocking of x with Cr$^{22+}$ and  broadband electron scattering (BES). Figure \ref{f:EES} considers the effects of  Fe$^{23+}$ and ESE. Figure \ref{fig:app} considers these processes in addition to Cr$^{22+}$ and broadband electron scattering (BES). 

The best model is shown in black and discussed above. That discussion also considered the Fe$^{23+}$ and ESE, which are shown with the red and blue lines. We found that x overlaps with a line of Cr XXIII, which produces additional RAD destruction. This was removed in the green line. Finally, broadband electron scattering traps some line radiation, which produces photoionization, and changes the ionization of the gas. This was removed in the magenta line. Together these processes account for the decrease in x in Figure \ref{fig:app}.

\section{New  CLOUDY commands}
Starting in CLOUDY version 17.03, ESE can be switched off using the command:
\hspace{2 mm}\begin{verbatim}
no scattering escape physics
\end{verbatim}
\hspace{2 mm} We exclude the effects of ESE in Figure \ref{f:EES} with this command.

In the presence of ESE, there will be a narrow and broad line component in each Gaussian in the Fe XXV K$\alpha$ complex. Only the narrow line components will be detected in the high-resolution telescopes.
The narrow components from each Gaussian can be extracted with CLOUDY using the command:
\hspace{2 mm} \begin{verbatim}
no scattering escape intensity
\end{verbatim}
\hspace{2 mm} in the input script. This is a change from the older version of CLOUDY that only had the \texttt{no scattering escape} command.


\bibliography{references}{}
\bibliographystyle{aasjournal}

\end{document}

%% file: macros.tex
\font\manual=manfnt at 7pt \def\dbend{\hbox{\raise0.9ex\hbox{\manual\char127\hspace{0.6em}}}}

\providecommand{\e}[1]{\ensuremath{\times 10^{#1}}}

\newcounter{INTERNALionstage}

\def\gtsim{\mathrel{\hbox{\rlap{\hbox{\lower4pt\hbox{$\sim$}}}\hbox{$>$}}}}
\def\lesssim{\mathrel{\hbox{\rlap{\hbox{\lower4pt\hbox{$\sim$}}}\hbox{$<$}}}}

\def\A{{\rm\thinspace \AA}}
\def\cm{{\rm\thinspace cm}}

\def\g{{\rm\thinspace g}}

\def\K{{\rm\thinspace K}}

\def\m{{\rm\thinspace m}}

\def\s{{\rm\thinspace s}}

\def\W{{\rm\thinspace W}}

\def\pscm{\mbox{$\cm^{-2}\,$}}

\def\pscm{\mbox{$\cm^{-2}\,$}}

\def\h0{\mbox{{\rm H}$^0$}}
\DeclareMathAlphabet{\vib}{OML}{cmm}{m}{it}
